\documentclass[a4paper,11pt]{article}
\usepackage{pos}

\title{LHC Experiments}

\author*[a,b]{Heather M. Gray}

\affiliation[a]{Department of Physics, University of California, Berkeley, \\
425 Physics South MC 7300 Berkeley, CA, 94720, USA}

\affiliation[b]{Physics Division, Lawrence Berkeley National Laboratory,\\
1 Cyclotron Road, Berkeley, CA 94720, USA}

\emailAdd{heather.gray@berkeley.edu}

\abstract{The field of experimental particle physics studies the fundamental particles and forces that constitute matter and radiation. Frequently the experimental tools used to enable this study are accelerators and detectors. The Large Hadron Collider (LHC) is the highest energy proton-proton accelerator currently operating and where the ATLAS and CMS collaboration discovered and are currently studying the properties of the Higgs boson. These notes provide a short introduction to accelerators and detectors using the LHC and its detectors as examples. The detector section will focus on two types of detectors extensively used today: tracking detectors and calorimeters. The notes will then discuss the algorithms used to process the information from the detectors and how that information is used for physics analysis using the search for the decay of the Higgs boson to bottom quarks. }

\tableofcontents

\begin{document}
\maketitle

  \section{Introduction}

\begin{figure}[hbtp!]
\centering
\vspace{-4mm}
\includegraphics[width=0.8\textwidth]{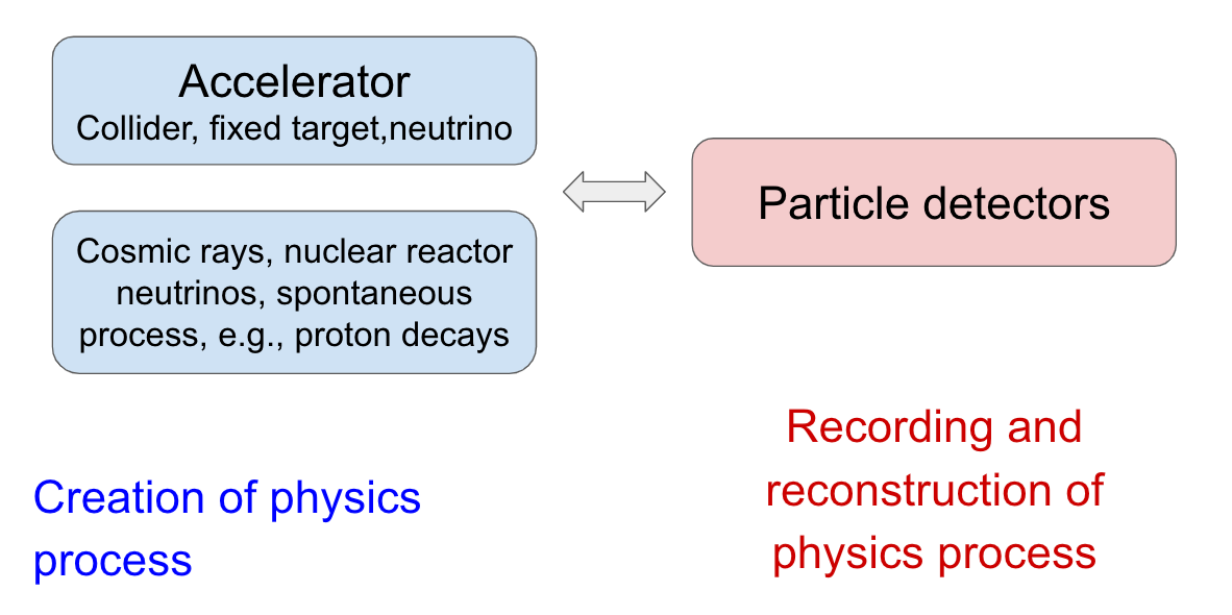}
\caption{
Illustration of the relationship between accelerators and detectors. where accelerators are used to create the physical process and detectors are used to record and reconstruct the physical process. From~\cite{shapiro}.
\label{fig:accdet}
}
\end{figure}

The field of experimental particle physics studies the fundamental particles and forces that constitute matter and radiation.  Studying these particles and forces, requires first creating the physical process of interest and then recording and reconstructing that physical process. Accelerators are generally used to create the physical process and particle detectors are used to record and reconstruct the physics process as illustrated in Figure~\ref{fig:accdet}.

The most appropriate method for production depends on the physics process being studied. Types of accelerators widely used today include colliders, fixed target, and neutrino beams. Natural processes such as cosmic rays, nuclear reactors, and other spontaneous processes can also be used to produce the particles instead of accelerators. Natural processes typically require fewer (or even zero) resources to produce particles, while accelerators provide greater control over the particles, direction, and production rate.

At colliders, new particles are created through high-energy collisions by exploiting $E = mc^2$ to transform the energy of the beams into new massive particles. Depending on the experimental needs, different particles are used in the beams. Currently, the most common choices are electron-positron and proton-(anti)-proton colliders. 
 
At fixed target experiments, particles (from man-made beams or not) are scattered on a target.  This can be used to study the structure of the target, as in Rutherford scattering experiments, or to study the structure of the beam particles. This can also be used to understand the interaction between the beam and the target, such as studying the weak current using neutrino-proton scattering. It can also be used to confirm the existence of the incoming particle, which is how neutrinos were observed and the strategy used by direct detection experiments to search for dark matter. This can also be used to measure a particle’s properties through direct interaction with experimental setup and measuring decay rates and kinematics. This has been used to understand the internal structure (spectroscopy), study the symmetries/properties of interactions (for example in searches for proton decay) or the confirmation of detailed Standard Model (SM) predictions (for example measurements of the muon magnetic moment).

These lecture notes aim to provide a brief introduction to selected topics in experimental particle physics using the experiments at the Large Hadron Collider (LHC)~\cite{Bruening:782076} as a particular example. Such a topic is typically covered by a full semester graduate lecture series, so, as such, here we are only able to provide an introduction to relevant topics. The interested reader is invited to consult the provided references for additional details.

First, we will introduce accelerators in general and the LHC in particular 
 in Section~\ref{sec:accelerators}). Next we discuss two of the key detector types used in particle experiments today: tracking detectors and calorimeters in Section~\ref{sec:detectors}. In Section~\ref{sec:det_physana} we will see how these individual components can be combined to form a full particle physics detector, discuss how the information from the detectors is processed using algorithms to obtain objects for physics analysis and illustrate how physics analysis is performed using a complex Higgs analysis.
  \section{Accelerators}
\label{sec:accelerators}
The key components of an accelerator include the beam, accelerating structures, and the magnets. The beam is a current of charged particles that are transported in an ultra-high vacuum and the particles are arranged in bunches with large numbers of particles per bunch. High multiplicity bunches are used because this increases the probability of individual particles interacting. The accelerating structures are used to accelerate the particles up to high energies. Most accelerators use either electric fields or radio-frequency waves for acceleration. A newer technique still under development is plasma wakefield acceleration~\cite{MangleLecture}, which has the promise to accelerate particles to high energies over much smaller distances. Magnets guide the beam in well-defined paths and focus the beams to small transverse areas. 

\begin{figure}[hbtp!]
\centering
\vspace{-4mm}
\includegraphics[width=0.9\textwidth]{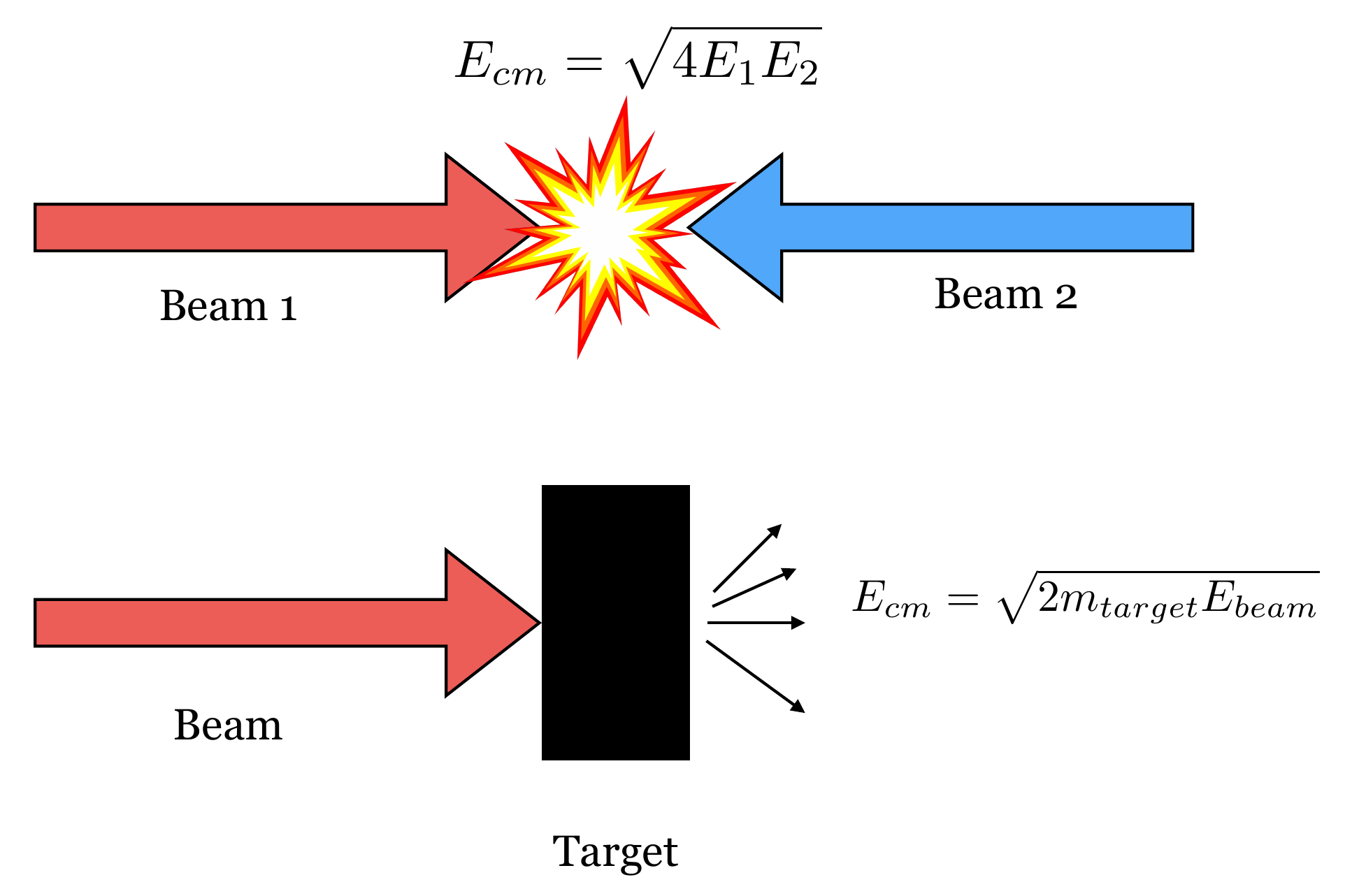}
\caption{
Illustration of two different types of particle accelerators. A collider, in which two beams collide, is illustrated at the top and a fixed target experiment, in which a single beam collides with a target, is illustrated at the bottom.
\label{fig:colliderfixedtarget}
}
\end{figure}

Two of the different types of accelerators are fixed target or accelerators. In colliders, where two beams collide, the center-of-mass energy of the collision is given by: $E_{cm} = 
\sqrt{4 E_1 E_2}$, where $E_1$ and $E_2$ are the energies of the two beams. All the energy of the beams is available for the hard scattering interaction and/or the creation of new particles. In fixed target collisions, where a single beams collides with a target, the center-of-mass energy is given by $E_{cm} = 
\sqrt{2 m_{target} E_{beam}}$, where $m_{target}$ is the mass of the target and $E_{beam}$ is the energy of the beam. While fixed target experiments have less energy available for producing new particles, a much wider variety of targets or beams can be used.

\subsection{Colliders}
At colliders the number of events is given by the following formula:

\begin{equation}
N_{event} = \sigma \mathcal{L} \Delta t
\end{equation}

where
\begin{itemize}
\item $\sigma$ is the cross section
\item $\mathcal{L}$ is the instantaneous luminosity
\item $\Delta t$ is the amount of time
\end{itemize}

The cross section is a measure of the probability of two particles interacting and the units are typically given in units of barns~\footnote{The barn is a unit of area and has colorful origins in the Manhattan Project~~\cite{barn}}. 
The units of the instantaneous luminosity are either $\mathrm{cm}^{-2}\mathrm{s}^{-1}$ or $\mathrm{pb}^{-1}\mathrm{s}^{-1}$. The instantaneous luminosity can be integrated over time to obtain the total luminosity, for example during a period of data taking. 

The luminosity is determined by the properties of the accelerator: 
\begin{equation}
\mathcal{L} = fn \frac{N_1 N_2}{4 \pi \sigma_x \sigma_y}
\end{equation}
where

\begin{itemize}
    \item f is the beam revolution frequency (LHC: 11 kHz)
    \item n is the number of bunches in the beam (LHC: typically 2808)
    \item N is the number of particles per bunch for each beam (LHC: $10^{11}$)
    \item $\sigma$ is the transverse size of the beam (LHC: 16 $\mu$m)
\end{itemize}
The goal of modern accelerators is to maximize the luminosity because this increases the number of events produced during any period of data taking. 

\subsubsection{Lepton vs Hadron Colliders}
In lepton colliders, the collisions are between two elementary, pointlike particles. This means that all the beam energy is available in the collision. Particles are predominantly produced at lepton colliders through the electroweak interaction.

In hadron colliders, the collisions are between the partons within two composite particles. This means that only a fraction of the beam energy is available in the collision, and this fraction is determined by the parton distribution function. Particles are predominantly produced at hadron colliders through the strong interaction. However, as will be discussed in the next section, circular proton-proton colliders can reach higher energies than electron-positron machines as they are less sensitive to synchrotron radiation.

Protons are produced by using a high electric field to ionize hydrogen atoms. Antiprotons are produced by shooting protons at a target material and filtering out antiprotons. Electrons are produced by heating metal until it spits off electrons. 

\subsubsection{Linear vs Circular Accelerators}
Accelerators are built in either linear or circular configurations. Circular accelerators have the advantage of being able to reuse the beam particles and the accelerator components. However, as the particles are continuously accelerating they lose energy through synchrotron radiation. The amount of radiation is given by the Larmor formula:

\begin{equation}
P = \frac{1}{6 \pi \epsilon_0} \frac{e^2 a^2}{c^3} \gamma^4
\end{equation}

which is inversely proportional to $m^4$ of the beam particles. 

The power radiated by different beam particles is
\begin{itemize}
    \item $W = 8.85 \times 10^{-5} \frac{E^4}{\rho}$  MeV per turn for electrons
    \item $W = 7.8 \times 10^{3} \frac{E^4}{\rho}$  keV per turn for protons
\end{itemize}

This means that electron-positron accelerators are strongly limited in the energies that they can reach due to synchrotron radiation. As protons are almost 2000 times heavier than electrons, they emit many orders of magnitude less synchrotron radiation. The idea of building a muon collider is often popular because they have the advantage of using an elementary particle as do electron-positron machines, but the mass of the muon means they emit significantly less synchrotron radiation.

\subsection{The Large Hadron Collider}

\begin{figure}[hbtp!]
\centering
\vspace{-4mm}
\includegraphics[width=0.9\textwidth]{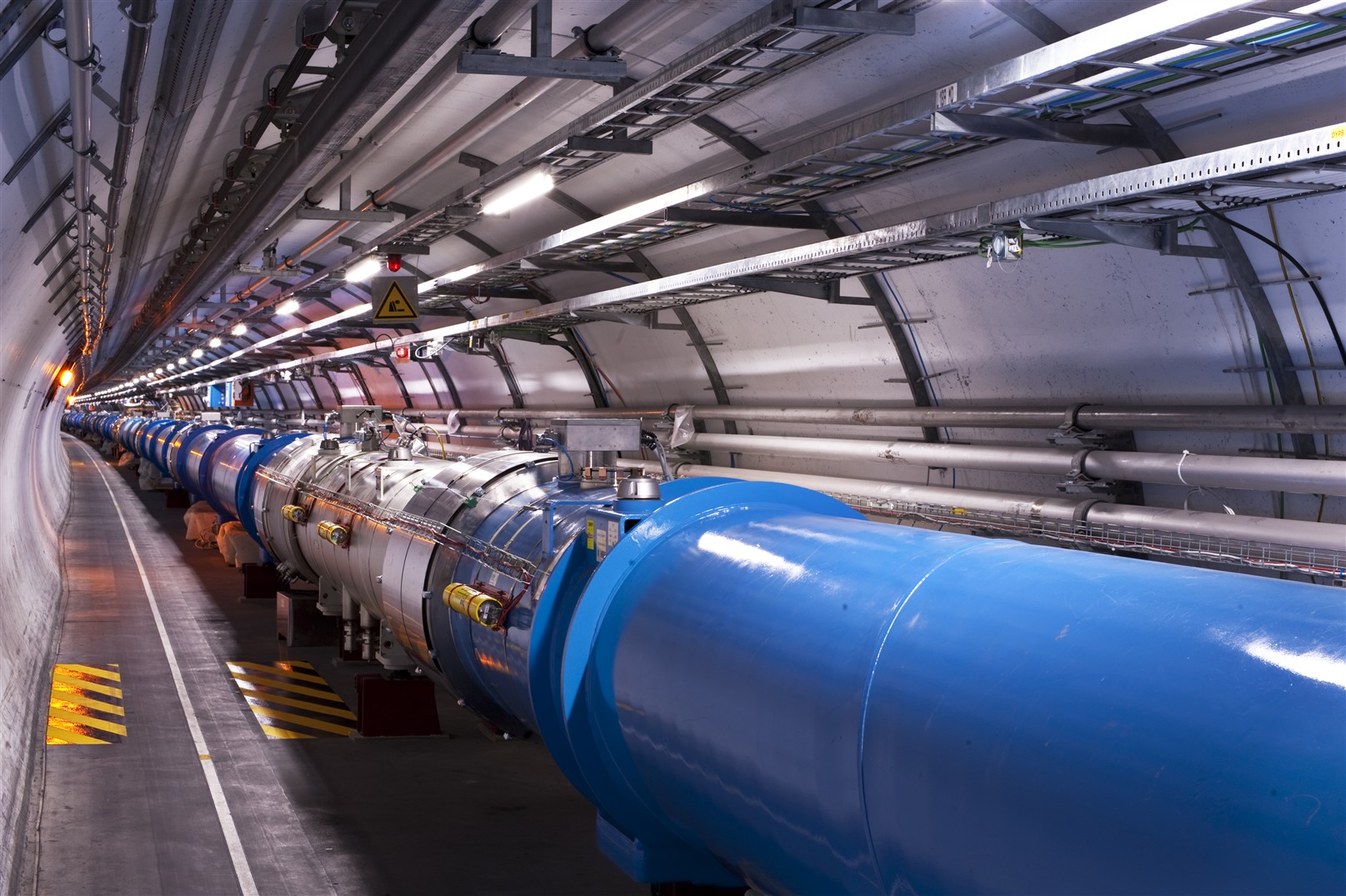}
\caption{
Dipole magnets, which are painted blue inside the LHC tunnel. The small yellow sign hanging from the ceiling indicates a oxygen deficiency risk in case of a gas leak. From Ref.\cite{Maximilien:1211045}
\label{fig:cernacc}
}
\end{figure}

The Large Hadron Collider (LHC) is a circular accelerator located just outside Geneva, Switzerland. It has a circumference of 26.7\,km, which translates into a curvature of 2.8\,km.  The design beam energy of 7\,TeV is determined by the radius of curvature and the field strength of the dipole magnets of 8.33\,T. There are 1276 dipoles within the LHC. They are each 15\,m long and made from the superconductor, Niobium Titanium (NbTi), which is operated at a temperature of 1.9\,K. There are two bores within each magnet, for the clockwise and anti-clockwise circulating beams. The current within the magnets is 11,800\,A. Additional magnets are used to control and focus the beam. The quadrupoles are oriented to either restore or anti-restore the focus of the beam. Higher-order magnets, up to decapoles, are used to provide more complex corrections.

\begin{equation}
\frac{1}{\rho} \approx 0.3 \frac{B}{p}
\end{equation}

Key parameters of the accelerator are the beam emittance, which determines how particles are confined to a small physical distance and a small range in momentum, and the amplitude, $\beta$. The amplitude is defined as:

\begin{equation}
\beta = \frac{\pi \sigma^2}{\epsilon}
\end{equation}
When the beam has low $\beta$, it is termed as squeezed and this translates into high luminosity for collisions. When the beam has high $\beta$ it is wide and straight. The special value of $\beta$ at the interaction point is termed $\beta^\star$. A challenge at the LHC is the fact that the high luminosity has a side effect of producing multiple proton-proton collisions each time the beams collide. These additional proton-proton collisions are known as pile up and during the second data-taking run of the LHC, the average was more than 40 pile up collisions per event. So far, there have been three data-taking runs for the LHC. The first from 2010-2012 reached a center-of-mass energy of 8\,TeV, the second from 2015-2018 reached a center-of-mass energy of 13\,TeV and the ongoing third run started in 2022 and has reached a center-of-mass energy of 13.6\,TeV.

\begin{figure}[hbtp!]
\centering
\vspace{-4mm}
\includegraphics[width=0.9\textwidth]{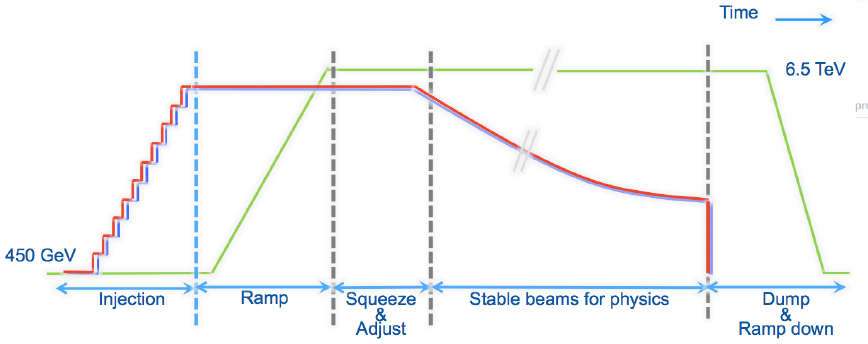}
\caption{
A typical fill from the LHC. The intensity of the two beams are protons are shown in red and blue. The energy of the beams is shown in green. From Ref.~\cite{RendeLecture1}
\label{fig:LHCLifecycle}
}
\end{figure}

Figure~\ref{fig:LHCLifecycle} shows a typical fill from the LHC. The beam moving clockwise around the LHC is typically known as the blue beam and the beam moving anticlockwise as the red beam. Bunches of protons are injected into the LHC in steps with an energy of 450\,GeV. Once all the bunches have been injected, the beams are accelerated to reach an energy of 6.5\,TeV.  Once the desired energy has been reached the beams are squeezed to minimize the crossing angle at the collision points and adjusted to obtain collisions. Stable beams are declared which allows the experiments to switch on their sensitive detectors and collect data for physics. Due to the collisions, the intensity of the beams decreases during this period. At the end of the data-taking the beams are dumped, the magnets are ramped down and the process is repeated for the next fill.

\subsection{Experiments at the LHC}

\begin{figure}[hbtp!]
\centering
\vspace{-4mm}
\includegraphics[width=0.9\textwidth]{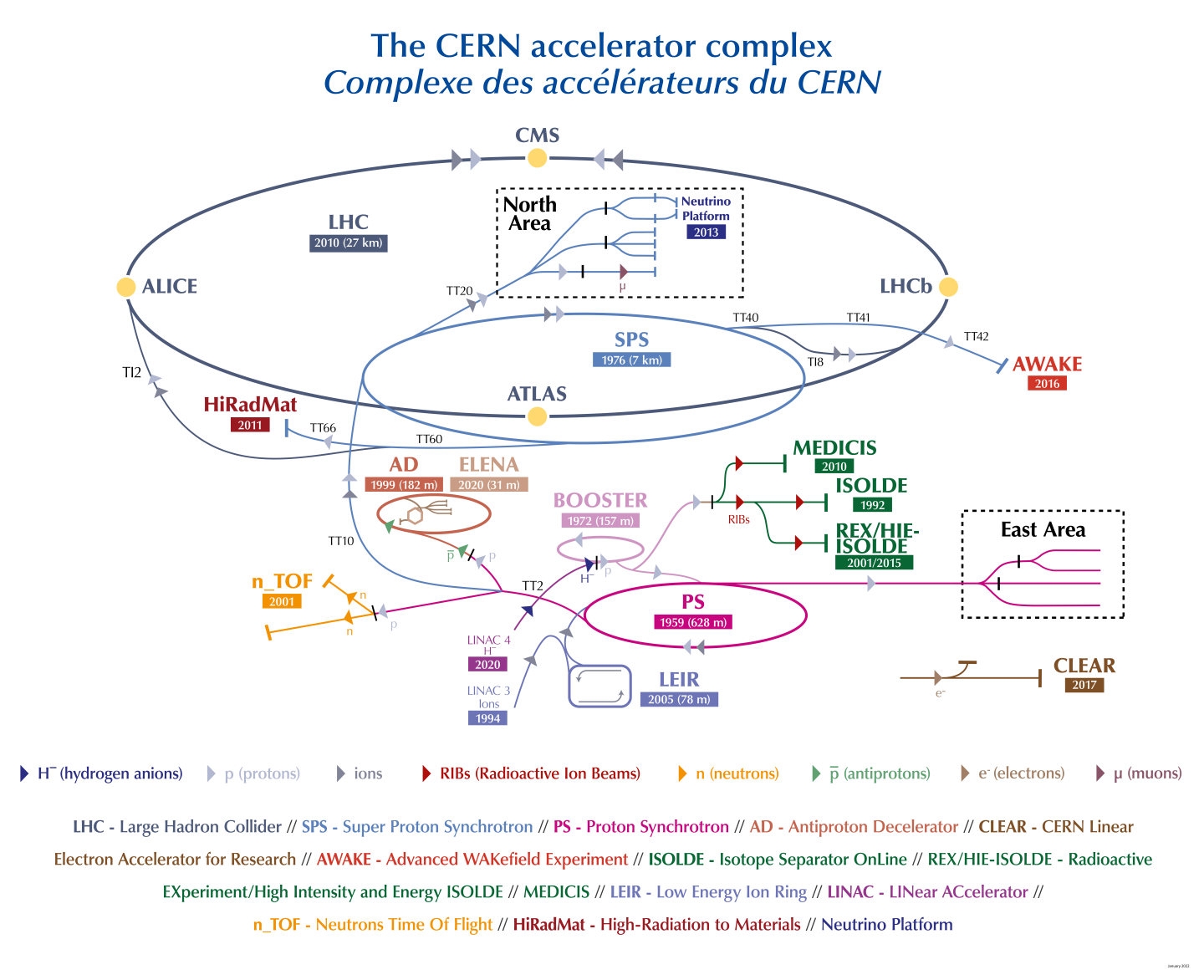}
\caption{
The CERN accelerator complex showing the different accelerators and detectors.From Ref.\cite{Lopienska:2800984}
\label{fig:cernacccomplex}
}
\end{figure}

There are four large and additional smaller physics experiments at different locations around the ring at the LHC as shown in Figure~\ref{fig:cernacccomplex}. The two general-purpose particle physics experiments are ATLAS~\cite{ATLAS_DetectorPaper} and CMS~\cite{CMS_DetectorPaper} and they are located at opposite sides of the ring. The ATLAS and CMS collaborations perform precision SM measurements, study the properties of the Higgs boson and search for physics beyond the Standard Model (BSM). ALICE~\cite{ALICE_DetectorPaper} is a dedicated heavy-ion experiment designed to study the properties of the quark-gluon plasma and LHCb~\cite{LHCb_DetectorPaper} is designed to explore the matter-antimatter asymmetry in the universe by studying b-physics, CP violation and other anomalies. Smaller experiments include TOTEM~\cite{TOTEM:2008lue} and LHCf~\cite{LHCf:2008lfy} which are designed to study forward physics and to measure the total cross section. MoEDAL-MAPP~\cite{MoEDAL:2014ttp} performs a dedicated search for magnetic monopoles and FASER~\cite{FASER:2022hcn} searches for new light particles and studies the properties of high energy neutrinos. Other proposed experiments include MATHUSLA~\cite{MATHUSLA:2018bqv,MATHUSLA:2020uve}, MilliQAN~\cite{Haas:2014dda} and CODEX-b~\cite{Aielli:2019ivi}. 

For additional introductory material about accelerators, please see References~\cite{EldredLectures,PrebysAcceleratorCourse,Edwards:1992unz, RendeLecture1,RendeLecture2} and Chapter 31, Accelerator Physics of Colliders in ~\cite{ParticleDataGroup:2022pth}.
  \section{Detectors}
\label{sec:detectors}
The particles produced in collisions are detected by exploiting the three relevant interactions in the Standard Model: electromagnetism and the weak and strong nuclear forces.

Particles with an appreciable lifetime are the focus for detection, i.e. stable particles, such that they are able to travel a macroscopic distance through the detector. Such stable particles are leptons, hadrons and photons. Different techniques are used to detect charged and neutral particles. 
Charged particles are detected through ionization. Photons convert into electron-position pairs, which are subsequently detected through ionization. Neutral leptons, i.e. neutrinos, are detected when they undergo a weak interaction and the particles produced through that weak interaction are detected. Charged or neutral hadrons undergo nuclear interactions and the secondaries produced in that interaction either interact with the detector material through the nuclear interaction or are detected through ionization. 

Due to the nature of quantum chromodynamics, quarks and gluons don't exist as free particles, but combine into hadrons. Other fundamental particles such as the weak vector bosons (W and Z) and the Higgs boson decay before they are able to travel any appreciable macroscopic distance so they are detected through the detection of their decay products.

Once a charged particle has been detected, its kinematic properties are measured to determine its four vector. The energy is measured by stopping the particle in the detector volume using either the electromagnetic or strong interaction and measuring the ionization energy. The absolute value of the momentum can be measured from the trajectory of a charged particle moving in a magnetic field. The direction of the momentum can be obtained by measuring the location of the interaction vertex, the direction of the trajectory or the location of the deposited energy. 

While a range of different techniques can be and are used to detect particles, we will focus here on two key types of detectors: tracking detectors to measure the trajectories of charged particles and calorimeters to measure the energy of charged and neutral particles.

\subsection{Tracking Detectors}
\label{sec:trackingdet}
\begin{figure}[hbtp!]
\centering
\vspace{-4mm}
\includegraphics[width=0.9\textwidth]{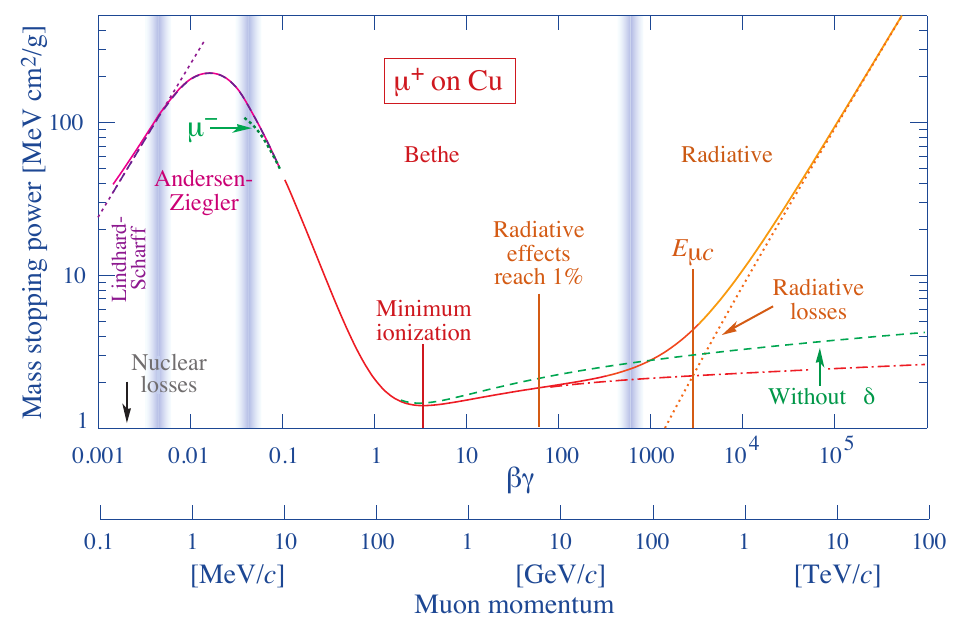}
\caption{
The energy loss of muons in copper as a function of the muon momentum. From Ref.\cite{ParticleDataGroup:2022pth}
\label{fig:bethebloch}
}
\end{figure}

Charged particles deposit energy in matter through the ionization of atoms. Key measures include the average ionization loss ($dE/dx$) and fluctuations in the amount of ionization energy deposited. Matter also affects charged particles through multiple scattering and bremsstrahlung. Bremsstrahlung is the radiation of photons from charged particles and is most important for electrons. Multiple scattering occurs when particles lose energy through many collisions with atoms and molecules, typically through Coulomb scattering. The energy lost by muons in matter as a function of the muon momentum is shown in Figure~\ref{fig:bethebloch}, showing how different energy loss processes dominate depending on the particle momentum Similar processes are observed for different particles and in different materials but over different energy ranges.  See Ref.~\cite{ParticleDataGroup:2022pth} for a detailed review of the energy loss of particles in matter.

Tracking detectors measure the properties of all charged particles. Commonly produced particles are $\pi^\pm$, $e^\pm$, $\mu^\pm$ and $K^\pm$. Tracking detectors measure the trajectory, momentum, and (for certain detectors) the particle type. Tracking detectors are typically located within a magnetic field, which allows the momentum to be determined from the curvature of the trajectory using the following formula: 
\begin{equation}
    \frac{p}{q} = B \rho
\end{equation}
Here $p$ is the momentum, $q$ is the charge, $B$ is the magnetic field, and $\rho$ is the radius curvature of the trajectory. This formula also shows that given a particular momentum of a particle, as the magnetic field increases, the curvature of the particle decreases, illustrating that magnetic field strength is a key parameter in tracking detector design.

Tracking detectors are also used to measure vertices of particle interactions. These vertices are used to identify the location of the collision between particles in the beams, known as the primary vertex, and the location of particle decays, known as secondary vertices.

Although historically, various experimental techniques have been used for tracking detectors, there are two main types used today: gaseous wire chambers and silicon detectors. 

\subsubsection{Multiwire Proportional Chamber}

\begin{figure}[h!]
\centering
\vspace{-4mm}
\includegraphics[width=0.4\textwidth]{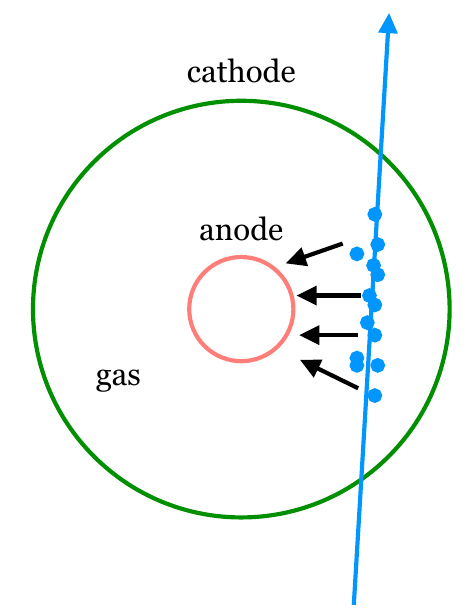}
\caption{
Illustration of a gaseous wire chamber. The cathode (green) is located on the outside and the anode (red) at the center. The ionization produced by a particle passing through the gas in the direction indicated by the arrow is shown in blue. 
\label{fig:wirechamber}
}
\end{figure}

A multiwire proportional chamber is a particular type of gaseous wire chamber detector and is illustrated in Figure~\ref{fig:wirechamber}. A charged particle moving upwards passes through the tube and emits primary electrons through ionization. The primary electrons drift toward the anode and are collected to measure the particle trajectory. In most cases these primary electrons are not sufficient to provide a reliable signal, therefore a high voltage is applied to accelerate the electrons and produce a detectable electronic signal. The ionization electrons take a certain amount of time to drift to the anode, which can be used to extract the time when the particle traversed the tracking detector. 

Multiwire drift chambers are similar to drift tubes but without individual tubes. Both planar and cylindrical geometries can be used and different arrangements of the cathode and anode wires can be used depending on the specific needs. They can easily cover large surface areas. For many years, multiwire proportional chambers (MWPC)~\cite{Charpak:1968kd} were widely used in particle physics experiments. The position resolution is determined by the wire spacing and is typically O(mm). Etched pads on the cathode can optionally be used to provide a measurement along the wire direction. The position resolution can be further improved by using the signal from multiple wires, but there is a fundamental limit due to the electrostatic repulsion. 

\begin{figure}[h!]
\centering
\vspace{-4mm}
\includegraphics[width=0.6\textwidth]{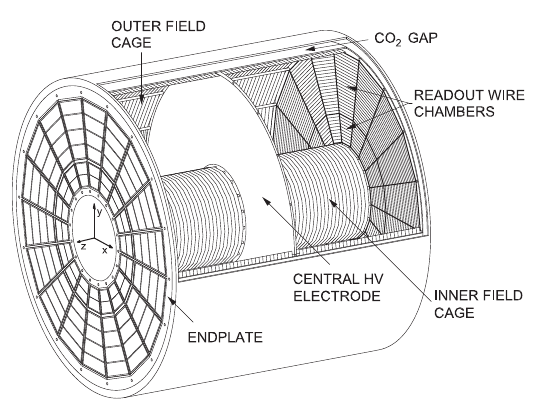}
\caption{
Schematic drawing of the ALICE TPC. The electrode is located at the center of the volume and the read out chambers are located on each end of the cylinder. From Ref.~\cite{Alme:2010ke} 
\label{fig:tpc}
}
\end{figure}

An example of a MWPC commonly in use today is a Time Projection Chamber (TPC), e.g. as used in the ALICE experiment. A schematic drawing of the ALICE TPC~\cite{Alme:2010ke} is shown in Figure~\ref{fig:tpc}. TPCs use a large volume of gas and a long drift distance. The electrode is located at the center of the cylinder and the signal is collected by the read out chambers at the end of the cylinder. Parallel electric and magnetic field are present in the chamber. The electric field ensures the electrons drift towards the read out chambers and the magnetic field minimizes the diffusion over the long drift distance. TPCs are excellent in minimizing the amount of material and provide good track resolution, however they cannot sustain high particle occupancies. 

\subsubsection{Silicon Detectors}

Silicon tracking detectors are very light and compact. They have excellent position resolution: for example pixel detectors typically have 10\,$\mu$m in precision and, as such, are typically the detector of choice at locations close to the collision point.

\begin{figure}[h!]
\centering
\vspace{-4mm}
\includegraphics[width=0.5\textwidth]{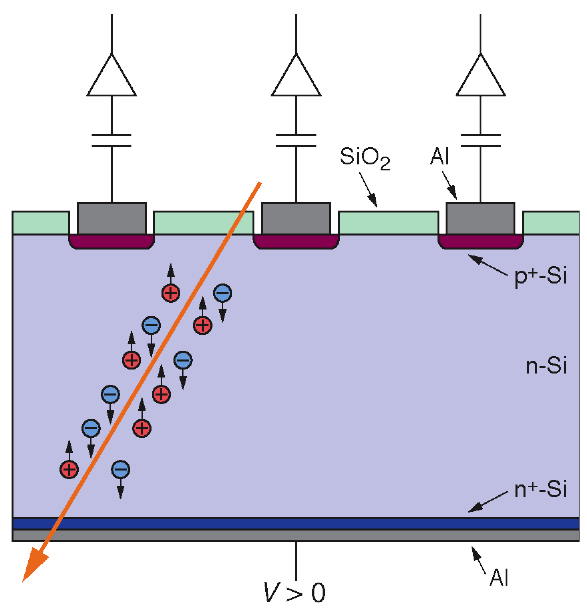}
\caption{Image of a typical n-type silicon strip detector. A particle passing through the depletion zone (orange arrow) produces electrons and holes. The charges drift to the electrodes and the signal is amplified by an amplifier. From Ref.~\cite{sidet}\label{fig:sidet}}
\end{figure}

A typical silicon detector, as shown in Fig.~\ref{fig:sidet}, is made by applying an inverse potential to a p-n junction in silicon. This produces a large region of depleted charge.  A particle passing through the depletion zones produces electrons and positrons. Due to the applied voltage, the charges drift to the electrodes and are read out. The signal is amplified by an amplifier which is connected to each of the strips.

The resolution of a silicon detector is dictated by the geometry. Typically silicon detectors either have small rectangles (known as pixels) or long strips. A typical pixel size in silicon detectors today is either 50 $\mu$m $\times$ 50 $\mu$m or 25 $\mu$m $\times$ 100 $\mu$m. Strips typically have a pitch of $70-80$ $\mu$m. Pairs of strips in strip detectors are often orientated with a small stereo angle with respect to each other, significantly improving the position resolution beyond the length of the strip.  

\begin{figure}[h!]
\centering
\vspace{-4mm}
\includegraphics[width=0.6\textwidth]{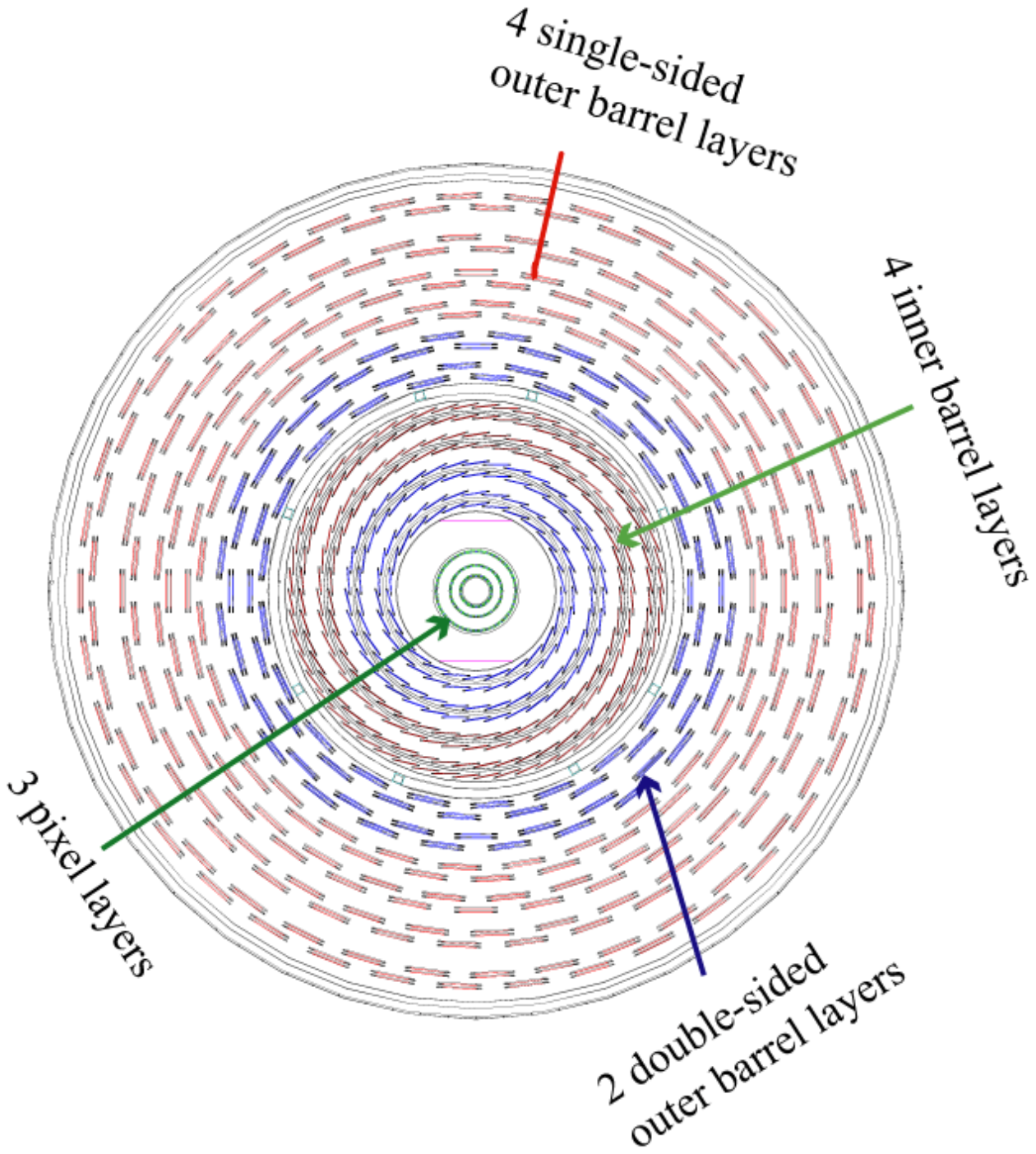}
\caption{Cross-sectional view of the original CMS Tracker perpendicular to the beam. The modules of the three layers of the pixel detector are shown in green. From Ref.~\cite{cms:tracker}\label{fig:cmstracker}}
\end{figure}

As an example, an image of the CMS tracking detector is shown in Fig.~\ref{fig:cmstracker}. The small rectangles indicate the positions of the individual modules. The silicon sensors are arranged in layers mounted on cylinders around the interaction point. The first three layers closest to the interaction point are the pixel layers and shown in green. These are surrounded by four inner barrel layers, two double-side outer barrel layers and four single-sided outer barrel layers. The CMS tracking detector is the world's largest silicon tracker. It has 200 m$^2$ of strip sensors and $11 \times 10^6$ read-out channels; as well as $\sim1$m$^2$ pixel sensors and $60 \times 10^6$ pixel channels. The precision of the CMS pixel detector is approximately 15$\mu$m in the transverse and longitudinal directions.

\begin{figure}[h!]
\centering
\vspace{-4mm}
\includegraphics[width=0.6\textwidth]{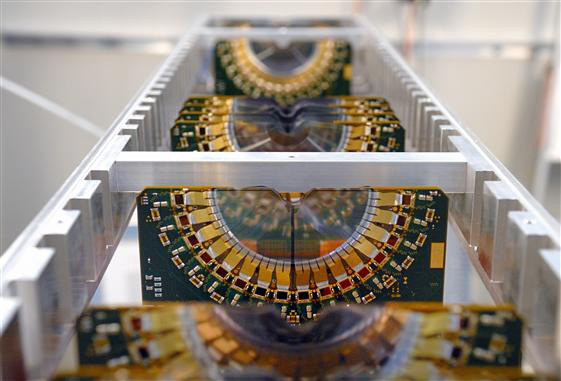}
\caption{
Photograph of the LHCb vertex detector, VELO. It contains 42 modules arranged in two rows of 21 double-side semi-circular silicon detector. Each semi-circle is approximately 8\,cm in diameter. From Ref.~\cite{lhcb:velo}
\label{fig:velo}
}
\end{figure}

An example of a silicon vertex detector is the LHCb Vertex Locator (VELO) which is shown in Fig.~\ref{fig:velo}. The VELO is located within the LHCb beam pipe, allowing the sensor to be as close to the collision point as possible. To ensure that the modules are protected during unstable beams before collisions, the VELO is designed to be able to move towards and away from the interaction point. The best hit resolution for the VELO is 4 $\mu$m at the optimal track angle~\cite{Aaij:2014zzy}.

\subsection{Calorimeters}
Calorimeters are (instrumented) blocks of matter that aim to measure the energy of incoming particles by stopping them in the material. Depending on the particle type they exploit either the electromagnetic or the nuclear interactions. The calorimeters are named correspondingly as electromagnetic or hadronic calorimeters.

\subsubsection{Showers in calorimeters}
\begin{figure}[h!]
\centering
\vspace{-4mm}
\includegraphics[width=0.8\textwidth]{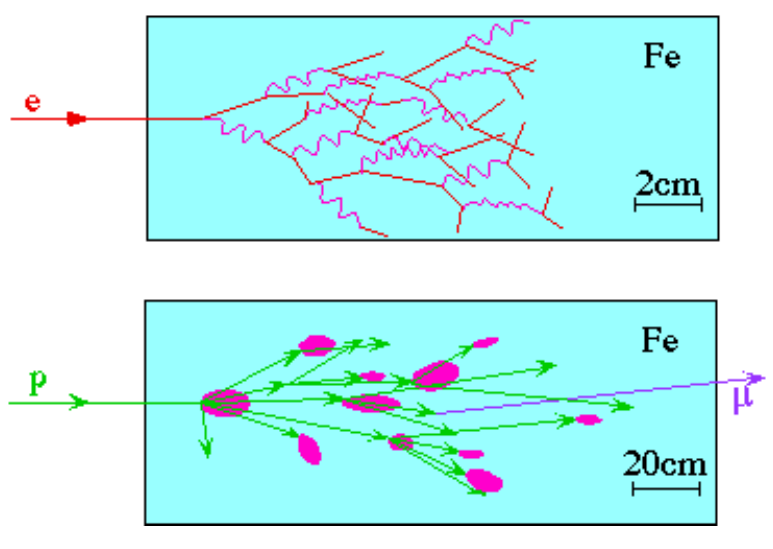}
\caption{
Illustration of the showers produced in iron by an electron (top) and a proton (bottom). Electrons are shown by red arrows; photons by pink curves, hadrons by green arrows and muons by purple arrows. The location of nuclear interactions are indicated by the pink ovals. From Ref.~\cite{BaileyLecture}
\label{fig:shower}
}
\end{figure}

 As the particles lose energy in the detector material, they produce showers of particles. Figure~\ref{fig:shower} illustrates typical showers for electrons and protons in iron. Both the types of particles produced and the distance over which they lose energy depend on the particle type.
 
\paragraph{Electromagnetic showers}

Photons and electrons produce electromagnetic showers. The showers they produce are essentially indistinguishable and consist of electrons and photons. 

Bremmstrahlung, in which radiation is produced by a charged particle, and pair production, in which an electron-positron pair is produced by a photon, dominate the shower development. The shower stops once the energy of the particles in the shower falls below the critical energy, which is approximately 10\,MeV. Below the critical energy, energy loss via ionization dominates for electrons and the Compton/photo-electric effect for photons. If the detector material is sufficiently thick such that it fully contains the shower, then the initial energy of the incoming particle is fully deposited. The signal is read out from the sensitive detector elements, which are often scintillators. The energy loss of electrons is characterized by the following formula:

\begin{equation}
- \left(
\frac{dE}{dx} \right)_{\mathrm{rad}} = \frac{E}{X_0}
\end{equation}

Here, $X_0$ is the radiation length which has units of g.cm$^{-2}$ and depends on the material.

The probability of a photon producing an electron-positron pair is given by the following formula

\begin{equation}
    \frac{d\omega}{dx} = \frac{1}{\lambda_{\mathrm{prod}}} e^{-x/\lambda_{\mathrm{prod}}} \lambda_{\mathrm{prod}} = \frac{9}{7} X_0
\end{equation}

Distances within calorimeters are expressed in terms of radiation lengths because this can abstract away the dependence on the particular material. For composite materials, $X_0$ is calculated with a weighted average of the radiation lengths of the individual materials. The weights are set by the relative mass fractions of the different materials.

The development of showers is characterized by their longitudinal and transverse sizes. In the longitudinal direction, the shower is characterized by the average number of particles at a depth $t$ expressed in radiation lengths. The number of particles in the shower increases exponentially with the depth and the energy of each particle decreases exponentially, i.e.\,
\begin{equation}
    N(t) = 2^t \Rightarrow E(t) = E_0 2^{-t}
\end{equation}

For accurate energy measurements a shower needs to be fully contained within a calorimeter, i.e.\,  all the particles produced in the shower need to be absorbed by the calorimeter rather than some of them escaping out the back. As the energy of the incoming particles increases, this means that logarithmically thicker calorimeters need to be used. The critical energy of a shower defines the required thickness of a calorimeter to fully contain the shower.

\begin{equation}
    E_c = E_0 2^{t_{\mathrm{max}}} \Rightarrow t_{\mathrm{max}} = \frac{log{E_0/E_c}}{log2}
\end{equation}

In the traverse direction, showers can be described by the Moli\`{e}re radius:

\begin{equation}
    R_M = X_0 ( 2 \mathrm{MeV}/E_c)
\end{equation}

The Moli\`{e}re radius gives the scale to contain the shower in the transverse direction and is used to determine the required granularity of the calorimeter. It also depends on the critical energy.

\paragraph{Hadronic Showers}
Similarly to electromagnetic showers, energetic hadrons cause the nuclei in the detector material to break up and form a cascade. These energetic hadrons are typically pions, neutrons and kaons. Particles are produced in the hadronic showers primarily through the strong interaction. The characteristic length scale of hadronic showers is the hadronic interaction length or nuclear interaction length, typically denoted as $\lambda_I$. This is the analog of the radiation length. The nuclear interaction length depends on the atomic number of the detector material and is approximated by the following formula:

\begin{equation}
    \lambda_I \approx 35 \mathrm{g/cm}^2 \mathrm{A}^{1/3}
\end{equation}

The energy deposited by hadrons is subject to significant fluctuations, much larger than for electromagnetic showers, which is why the energy resolution of hadronic calorimeters is worse than electromagnetic calorimeters.

\subsubsection{Types of Calorimeters}

Calorimeters are also either total absorption or sampling calorimeters. Passive material in calorimeters is used to produce showers, while active material records the particles produced in showers.

\paragraph{Total Absorption Calorimeters}
In total absorption calorimeters, particles stop, i.e. lose all their energy, within the active material. This means that the signal is directly proportional to the particle energy and the energy resolution typically scales as $\approx 1/E^{1/4}$ with the energy.

\begin{figure}[h!]
\centering
\includegraphics[width=0.8\textwidth]{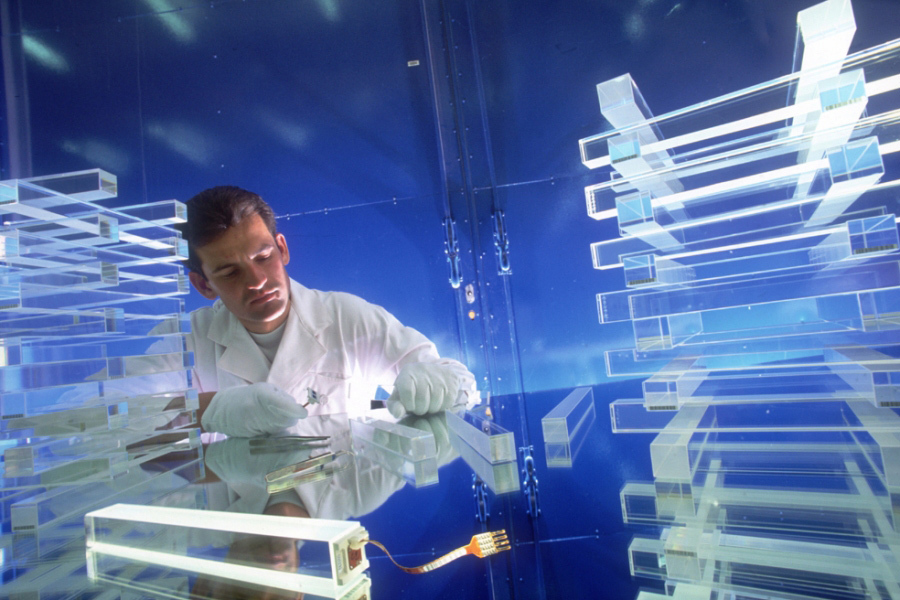}
\caption{
Tests of the lead-tungstate crystals of the CMS calorimeter during construction. From Ref.~\cite{Brice:1431477}
\label{fig:cmscrystals}
}
\end{figure}

The CMS electromagnetic calorimeter~\cite{CMS:1997ema} is an example of a total absorption calorimeter. It is made from lead tungstate (PbW0$_4$) crystals as shown in Fig.~\ref{fig:cmscrystals}. The dimensions of each crystal are 22 $\times$ 22 $\times$ 230 mm. Ionization from the incoming particle is absorbed in the crystals and light is emitted through scintillation. Photodetectors collect and amplify the light. The resolution of the CMS calorimeter is O(1\%) for 30\,GeV particles. 

\paragraph{Sampling Calorimeters}
Sampling calorimeters rely on a high $Z$ passive material to induce the shower and an active material to partially record the shower via ionization or scintillation. They typically have several layers of both passive and active material and each layer can be optionally segmented. The total energy is calculated after correcting for the sampling fraction, which is defined as follows
\begin{equation}
f = \sum \frac{dE}{dx}_{\mathrm{active}}/\sum \frac{dE}{dx}_{\mathrm{absorber}}
\end{equation}

In sampling calorimeters the energy resolution is given by the following formula
\begin{equation}
    \frac{\sigma_E}{E} = \frac{A}{\sqrt{E}} \oplus \frac{B}{E} \oplus C
\end{equation}

The three constants in the equation are:
\begin{itemize}
    \item A: stochastic term
    \item B: noise term
    \item C: constant term
\end{itemize}
The stochastic term depends on fluctuations in the shower development; the noise term on the amount of noise in the detector; and the constant term on the calibration process.    

The ATLAS electromagnetic calorimeter~\cite{ATLAS:1996guk} is an example of a sampling calorimeter. In the barrel region, $|\eta| < 1.475$, lead is used as the absorber and liquid argon as the scintillating material. A special accordion structure with a honeycomb pattern is used to ensure that there are no gaps between the layers along the azimuthal direction where particles would be missed. The total thickness of the calorimeter is at least 22 radiation lengths in the barrel and 24 radiation lengths in the end-cap.  The ATLAS calorimeter is also segmented into layers in the direction moving radially from the interaction point, i.e. the direction of the longitudinal development of the shower. It typically includes a fine-grained presampling layer and up to three additional layers. The sizes of the cells in the calorimeter vary according to their location in the detector and typically are $\delta \eta = 0.025$ and $\delta \phi = 0.0245$.

\begin{figure}[h!]
\centering
\includegraphics[width=0.8\textwidth]{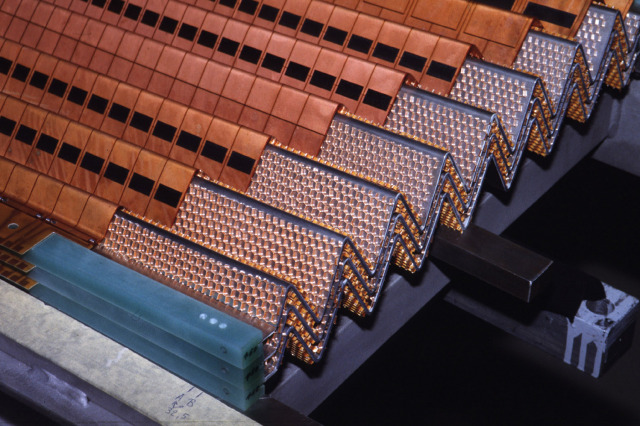}
\caption{
A slice of the ATLAS liquid argon barrel calorimeter showing the accordion structure. From Ref.~\cite{CERN-EX-9308048_09}
\label{fig:atlascalo}
}
\end{figure}

For additional introductory material about detectors, please see the following references ~\cite{BortolettoLecture1, BortolettoLecture2} as well as Chapter 34, Passage of particles through matter and Chapter 35, Particle detectors at accelerators in ~\cite{ParticleDataGroup:2022pth}. For material about tracking detectors, please see ~\cite{BerryLectures} and ~\cite{KolbergLectures}.
   \section{From Detectors to Physics Analysis}
\label{sec:det_physana}

\subsection{Full Detectors}
Full detectors for particle physics experiments are constructed by combining the detector elements discussed in Section~\ref{sec:detectors} to ensure that the particles produced in the collision are recorded. Different detector layouts are used depending on the physics goals of the experiment coupled with space and financial constraints. Typical geometries include cylindrical detectors with wheels at either side known as end caps or large wedges. Cylindrical detectors have the advantage of detecting particles at almost all angles produced in collisions, while the wedge-shaped detectors are easier to construct and maintain, typically less expensive, and can detect particles produced at angles very close to the direction of the beam. 

\begin{figure}[hbtp!]
\centering
\vspace{-4mm}
\includegraphics[width=0.9\textwidth]{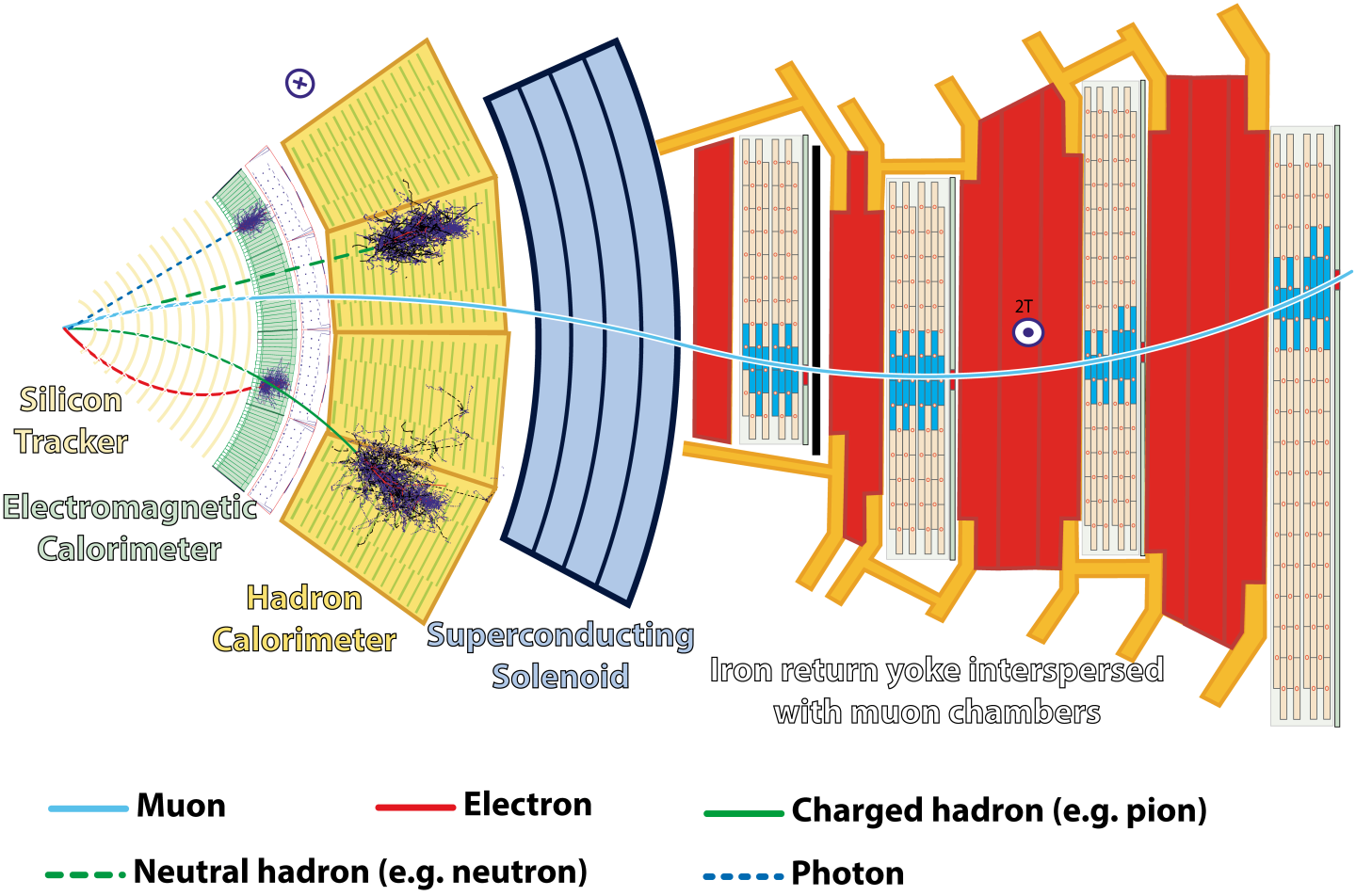}
\caption{
View of a wedge of the CMS detector perpendicular to the direction of the beam. The interaction point where the protons collide is located on the left. Moving from the interaction point to the right (which would be radially outwards), the silicon tracker, electromagnetic calorimeter, the solenoidal magnet and the muon detector. Different particles are indicated by curves, with solid curves for charged particle and dashed curves for neutral particles which do not interact in the tracking detector. The showers produced by the calorimeters are indicated in black. From Ref.\cite{Barney:2120661}
\label{fig:cmsslice}
}
\end{figure}

Fig.~\ref{fig:cmsslice} shows a slice of the CMS detector, an example of a cylindrical detector, in the direction perpendicular to the beam. The interaction point, where the two beams from the LHC collide, is located on the left and moving rightwards moves radially away from the interaction point~\footnote{The coordinate system used by the LHC experiments is a right-handed coordinate system with the $x$-axis pointing towards the center of the LHC tunnel, and the $z$-axis along the tunnel}. The silicon tracking detector is located closest to the interaction point (yellow layers) and the trajectories of charged particles that it measures are shown with solid lines. The silicon detector is located as close to the interaction point as possible to ensure that the charged particles pass through as little passive material as possible to minimize distortions to their trajectories. The electromagnetic calorimeter, located outside the tracking detector, is shown in green. The showers produced by electrons and photons in the electromagnetic calorimeter are shown in black. The hadronic calorimeter is located outside the electromagnetic calorimeter and showers of protons and neutrons are shown in black. The electromagnetic calorimeter is located closer to the interaction point than the hadronic calorimeter to minimize the impact of passive material on electromagnetic showers, because in most cases the energy resolution of electromagnetic particles needs to be more precise than that of hadronic particles. The tracking detectors and calorimeters are placed within a superconducting solenoid to curve their trajectories to enable the measurement of their trajectories. Muon chambers are located outside the superconducting solenoid and are shown in red. The iron return yoke for the magnet is interspersed between the muon chambers and bends the particles in the opposite direction so that the momenta of muons can be measured while passing through the muon chambers. Muon detectors are usually positioned as the outermost detectors because their trajectories are only minimally distorted by the detector material and this decreases the background from other particles punching through the calorimeter.

\begin{figure}[hbtp!]
\centering
\vspace{-4mm}
\includegraphics[width=0.9\textwidth]{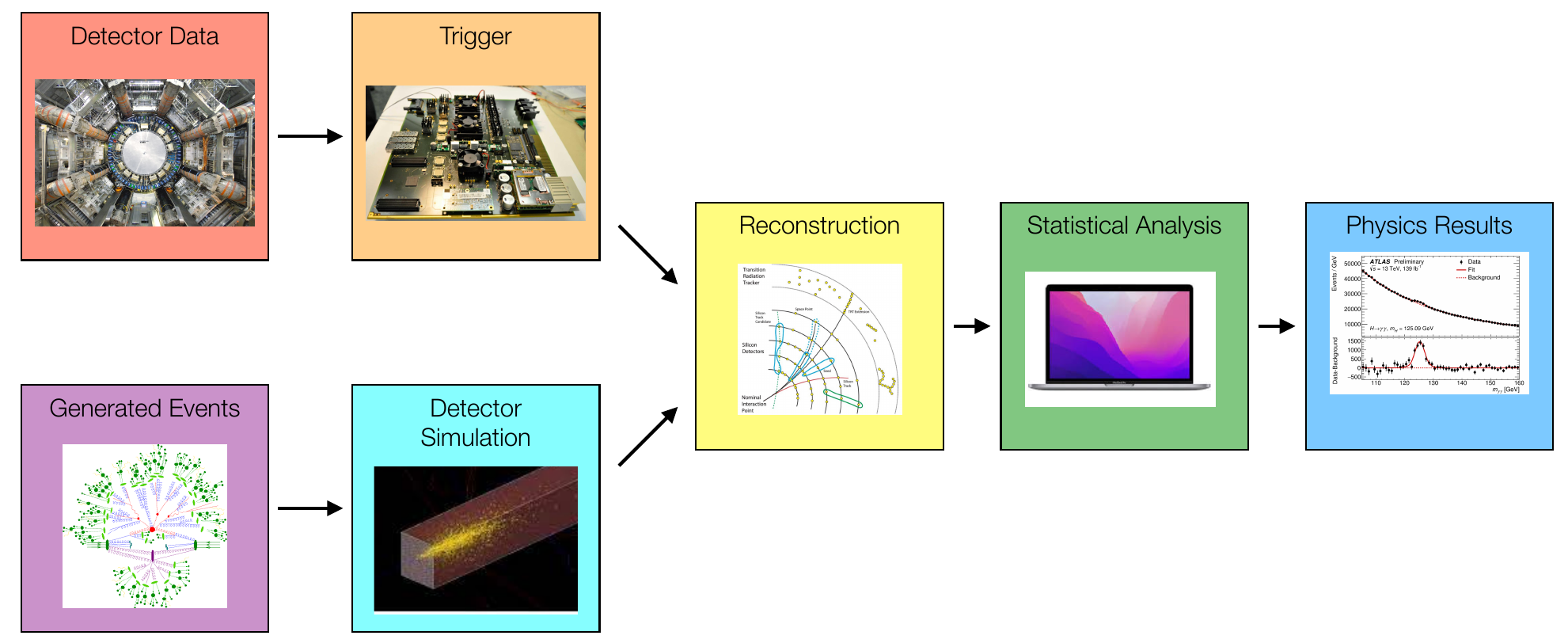}
\caption{
Illustration of the processing steps for the data (top) and simulation (bottom) from the detector or generation until physics analysis. 
\label{fig:detphys}
}
\end{figure}

The raw output from the read-out electronics of the detectors needs to be processed by sophisticated algorithms to obtain the reconstructed objects used in physics analyses. The steps in the processing chain are illustrated in Figure~\ref{fig:detphys}. Data processing begins with raw detector output where the events to be reconstructed are selected using a hardware trigger. Events are reconstructed and then analyzed before passing through a statistical procedure to obtain the final physics result. In almost all analyses simulated data is needed to perform the physics analysis. The simulated data is produced using event generations and then passed through a simulation of the detector response including the trigger. The simulated data is then processed in as similar a way as possible to the experimental data to minimize the impact of systematic uncertainties.

A large number and wide variety of physics analyses are performed at general-purpose particle physics detectors such as ATLAS and CMS.  Both experiments have published over a thousand papers over the 13 years since data-taking started. This section will provide examples of how the data from the different detectors is reconstructed to obtain the physics objects and then illustrate how these can be used in analysis with a brief overview of a complex analysis using that data. This section will not be exhaustive; key topics not covered here are electron, muon, $\tau$, and missing energy reconstruction. In addition, the trigger, which is used to select events is not discussed.

\subsection{Reconstruction of Physics Objects}
\subsubsection{Tracking}
As discussed in Section~\ref{sec:trackingdet}, tracking detectors consist of an array of tracking elements arranged in layers. They are designed to record the trajectories of charged particles including charged hadrons and muons by recording a series of space point measurements. Tracking algorithms aim to identify the set of measurements (known as hits) corresponding to a single charged particle and then extract the parameters of the trajectory by fitting the set of measurements. The fit assumes a (locally) helicoidal trajectory and typically uses a $\chi^2$ fit. The outputs of the track fit are the momentum and the charge of the track. Modern approaches to track finding and fitting typically rely on the Kalman Filter although approaches from machine learning are currently being explored.

\begin{figure}[h!]
\centering
\vspace{-4mm}
\includegraphics[width=0.8\textwidth]{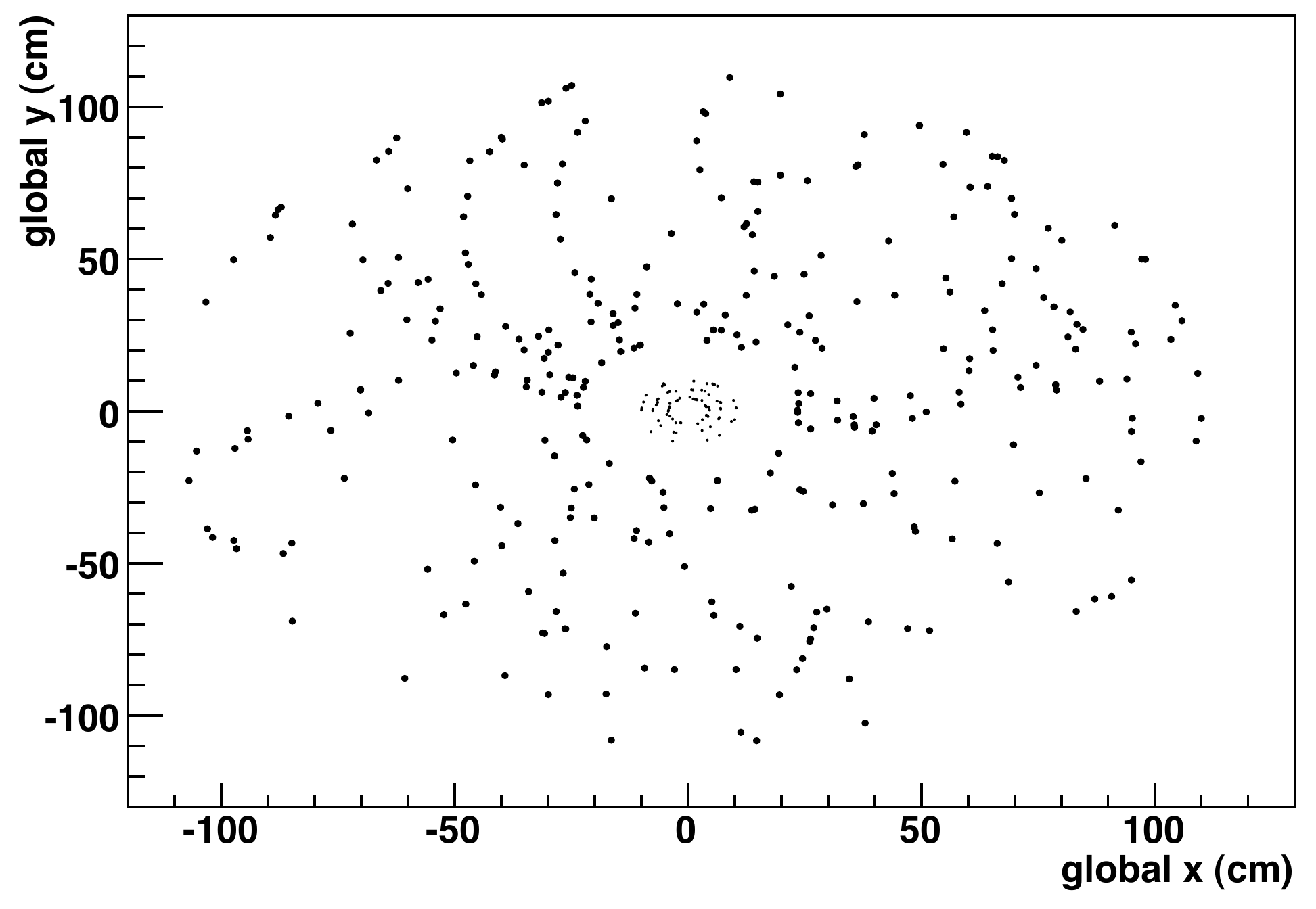}
\caption{
Example of a low multiplicity event containing a 50\,GeV track. Can you identify the hits corresponding to the track by eye? From Ref.~\cite{Dominguez}.
\label{fig:findtrack}
}
\end{figure}

\begin{figure}[h!]
\centering
\vspace{-4mm}
\includegraphics[width=0.8\textwidth]{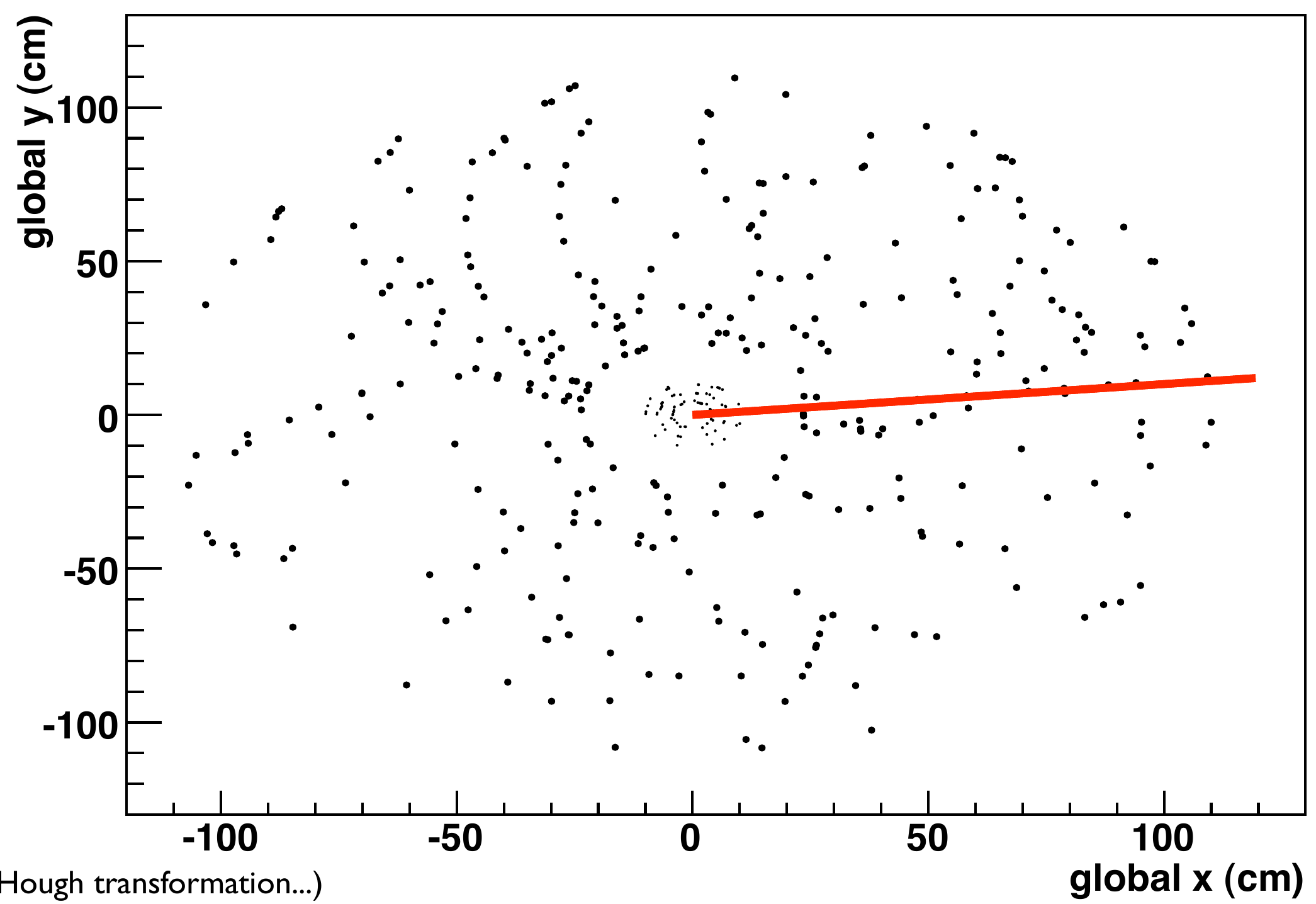}
\caption{
Example of a low multiplicity event containing a 50\,GeV track indicated in red. From Ref.~\cite{Dominguez}.
\label{fig:findtracksoln}
}
\end{figure}

Figure~\ref{fig:findtrack} shows an example of the hits in a tracking detector for an event containing a single high momentum (50\, GeV) track and a number of lower momentum tracks. Figure~\ref{fig:findtracksoln} shows the same event with the 50\, GeV track indicated in red. At the LHC, there are typically 500 tracks with momentum above 500\, MeV in each recorded event, but most come from pile up events.

Tracking algorithms are typically highly efficient, e.g. if a particle passing through the detector deposits energy in all the layers, the reconstruction efficiency is approximately 100\%. They also have a tiny fake rate, which can be as low as $10{-4}$ fakes compared to the total number of reconstructed tracks~\cite{ATLAS:2021yvc,ATLAS:2016nnj,CMS:2014pgm}. The performance is limited by the rate of material interactions and hence the amount of detector material. In the cores of dense jets, the performance is limited by the size of the detector elements as particles are produced very close together.

\clearpage

\subsubsection{Vertexing}
Tracks of charged particles are used to determine the location where physical processes occurred. This location is termed a vertex and it lies at the interaction of multiple tracks. There are two different types of vertices. The primary vertex aims to identify the interaction point(s) where the collision occurred. The secondary vertex aims to identify the decay position of particles with appreciable lifetimes, including bottom and charm quarks. 

Typical displacements of primary vertices from the center of the detector are comparable in scale to the resolution of the sensitive silicon detector elements. This means that precise measurements from detectors near the interaction point are required. 

Vertex reconstruction algorithms use the outputs of the track reconstruction algorithms as inputs. The algorithms follow an iterative procedure to identify the groups of tracks corresponding to a particular vertex. Kalman Filter algorithms are often used to determine the position of the vertex~\cite{ATLAS:2016nnj,CMS:2014pgm}.

At the LHC, a key challenge for vertexing is the amount of pile up. During run 3 there have been, on average, more than 40 additional interactions in each event. These additional interactions can cause the efficiency for the vertex reconstruction algorithm to decrease, more fake vertices to be reconstructed and vertices to be split or merged. The vertex reconstruction algorithms can also be used to mitigate the impact of pile up on other reconstructed objects by identifying those particles which originated from other vertices.

\subsubsection{Jets}
Due to the nature of the strong interaction, quarks and gluons appear experimentally as sprays of particles. Jet algorithms are used to reconstruct the properties of the quarks and gluons by grouping together particles. There are two general categories of jet algorithm. Geometric cluster algorithms group particles based on their angular separation. Recombination cluster algorithms group particles using momentum-based metrics. Examples of the latter are the $k_T$~\cite{Catani:1993hr, Ellis:1993tq} and anti-$k_T$~\cite{Cacciari:2008gp} algorithms, which rely on the following distance metrics:
\begin{itemize}
    \item $k_T$ algorithm: $d_{ij} = min(k_{Ti}^2, k_{Tj}^2) \frac{\Delta R_{ij}}{R}$ 
    \item anti-$k_T$ algorithm: $d_{ij} = min(k_{Ti}^{-2}, k_{Tj}^{-2}) \frac{\Delta R_{ij}}{R}$ 
\end{itemize}
where $k_T$ is the momentum and $R_{ij}$ is the angular separation between pairs of particles.

The anti-$k_T$ algorithm has many nice experimental and theoretical properties, which have led it to be adopted as the algorithm of choice at the LHC. For example, jets dominated by low transverse momentum particles produced in soft fragmentation tend to have a conical shape and the algorithm has nice theoretical properties including infrared and collinear safety. Infrared safe algorithms yield the same set of jets after modifying the event by adding soft radiation. Collinear safe algorithms yield the same set of jets after introducing a collinear splitting of one of the inputs.

Jet algorithms are typically very efficient and the challenge lies in optimizing the jet energy scale and resolution. Detailed calibration procedures are required before the jets can be used for physics analysis. The uncertainty on the jet energy scale is typically 1\% for intermediate momentum jets in the center of the detector and increases to a few percent at lower and higher energies.  The energy resolution varies from $\approx$ 25\% at low energy to 6\% at high energy~\cite{CMS:2016lmd,ATLAS:2020cli}.

Jets containing bottom ($b$) and charm ($c$) quarks can be identified by identifying the secondary vertices produced in their decays within jets. Current approaches for jet flavor tagging rely extensively on machine learning including Graph Neural Networks. The jets identified by their jet flavor tagging algorithms are known as $b$-tagged and $c$-tagged jets. The performance of flavor tagging algorithms always requires trading off between the efficiency of the targeted flavor and degree of background rejection. This is because the output of the algorithm typically produces a discriminant where certain values characterize more signal-like jets and other values characterize more background-like jets. A cut is then placed on the discriminant to yield a particular working point of the algorithm.

For example, with a $b$-tagging efficiency of 70\% , the $c$-jet background is reduced by an order of magnitude and the light-jet background by a factor of 600. Tagging $c$-jets is more difficult and with a $c$-tagging efficiency of 30\%, the $b$-background is reduced by almost an order of magnitude and the light jet background by a factor of 70~\cite{ATLAS:2022qxm,CMS:2017wtu}.

\subsection{Physics Analysis}
The final step in producing a physics result is the analysis. Analyses use the physics objects from the reconstructed and simulated data as input and perform a statistical analysis. Depending on the scale of an analysis there are often multiple processing steps, e.g. producing smaller files that can be analyzed more quickly, and each analysis workflow is run many times.

\subsubsection{Higgs coupling to bottom quarks}

As an example, we will consider the measurement of the coupling of the Higgs boson to bottom quarks. After the discovery of the Higgs boson at the LHC in 2012~\cite{Aad:2012tfa, Chatrchyan:2012xdj}, many of its properties including the mass, width, parity and coupling to other SM particles have been studied in dedicated analyses~\cite{ATLAS:2022vkf,CMS:2022dwd}. Higgs analyses vary widely in terms of the physics objects used as well as the analysis techniques used.

The fraction of all decays of a particle to a particular set of particles is known as the branching ratio. The Higgs coupling to bottom quarks has the largest branching ratio of all Higgs decays and can be used to probe the strength of the Higgs coupling to fermions and quarks. The so-called $H \rightarrow b \bar{b}$ analysis is performed using associated production of a Higgs boson with a weak vector boson, because this reduces the background from pairs of $b$-jets by many orders of magnitude~\footnote{This is because the cross section to production a $W$ boson and a pair of $b$-jets is much lower than to produce a pair of $b$-jets.} and the decay of the weak vector boson provides and an efficient trigger. Nonetheless, it is a very challenging analysis as the resolution of jets is much worse than for leptons and the analysis has large and varied backgrounds.
 
The analysis uses information from all detectors and almost all reconstructed physics objects. Techniques used in the $H \rightarrow b \bar{b}$ analysis which are commonly used in many Higgs analyses are categories, machine learning and a profile likelihood fit as will be further discussed later.

The analysis is performed separately for three channels differentiated by the number of leptons produced in the decay of the vector boson:

\begin{itemize}
    \item $W \rightarrow \ell \nu$: 1 lepton
    \item $Z \rightarrow \ell \ell$: 2 lepton
    \item $Z \rightarrow \nu \nu$: 0 lepton
\end{itemize}

In all cases, only electrons and muons are used as the `leptons' because they have much better reconstruction efficiency and resolution than $\tau$ leptons. 

A basic event selection is applied to identify events containing two $b$-tagged jets and the appropriate number of leptons and neutrinos for the channel. The analysis has a large number of significant backgrounds, which are not necessarily well-modeled by the Monte Carlo generators used to produce the simulated data. Some of the largest backgrounds include events containing pairs of top quarks, event containing weak vector bosons and jets (particularly when the jets are heavy-flavor jets) and events containing pairs of weak vector bosons. This is because those backgrounds results in a very similar set of final state particles to the signal. For example, in the 1-lepton channel, the signal events have one lepton, missing energy and two $b$-jets. Events containing pairs of top quarks almost always decays to two $W$ bosons and two $b$-jets. This means that the background can mimic the signal when one of the $W$ bosons decays leptonically to a lepton and missing energy and the other decays hadronically to a pair of jets. The main difference between the background and the signal is the presence of the additional two jets.

After event selection, the invariant mass of the two b-jets can be fit to identify the Higgs signal and the background. The sensitivity of the analysis is improved, however, by using additional kinematic variables which have different properties for signal and background to separate events into multiple categories. The ratio of the signal to the background varies between the categories, which increases the sensitivity of the analysis.

These variables are also combined into a discriminant using machine learning techniques, which has better separation between signal and background than the mass, hence improving the sensitivity of the analysis. The three most important variables for the analysis are the transverse momentum of the vector boson (which is the same as the transverse momentum of the Higgs boson, but measured with better experimental precision), the invariant mass of the pair of b-jets and the opening angle between the b-jets. For example, the Higgs boson has a much harder transverse momentum distribution than the backgrounds. In addition, events are split into categories according to how the number of signal and background events vary. These categories are used to extract the signal but also as control regions. 

A profile likelihood fit to the final discriminating distribution is used to extract the signal and constrain the backgrounds. The likelihood depends on the parameter of interest, $\mu$, which is a scale factor applied to the strength of the coupling of the Higgs boson to bottom quarks, and the nuisance parameters which describe the systematic uncertainties, $\vec{\theta}$. The fit is constructed such that $\mu = 1$ corresponds to the SM predictions. Values of $\mu$ less than one correspond to weaker coupling strengths of the Higgs to bottom quarks than predicted by the SM and values of $\mu$ greater than one correspond to stronger couplings strengths than predicted by the SM. The profile likelihood technique is powerful and allows many different regions and uncertainties to be included. There can be hundreds or even thousands of nuisance parameters in a typical fit. The technique also uses the data to constrain the size of the nuisance parameters, which reduces the final uncertainty on the measurement.

\begin{figure}[hbtp!]
\centering
\vspace{-4mm}
\includegraphics[width=0.9\textwidth]{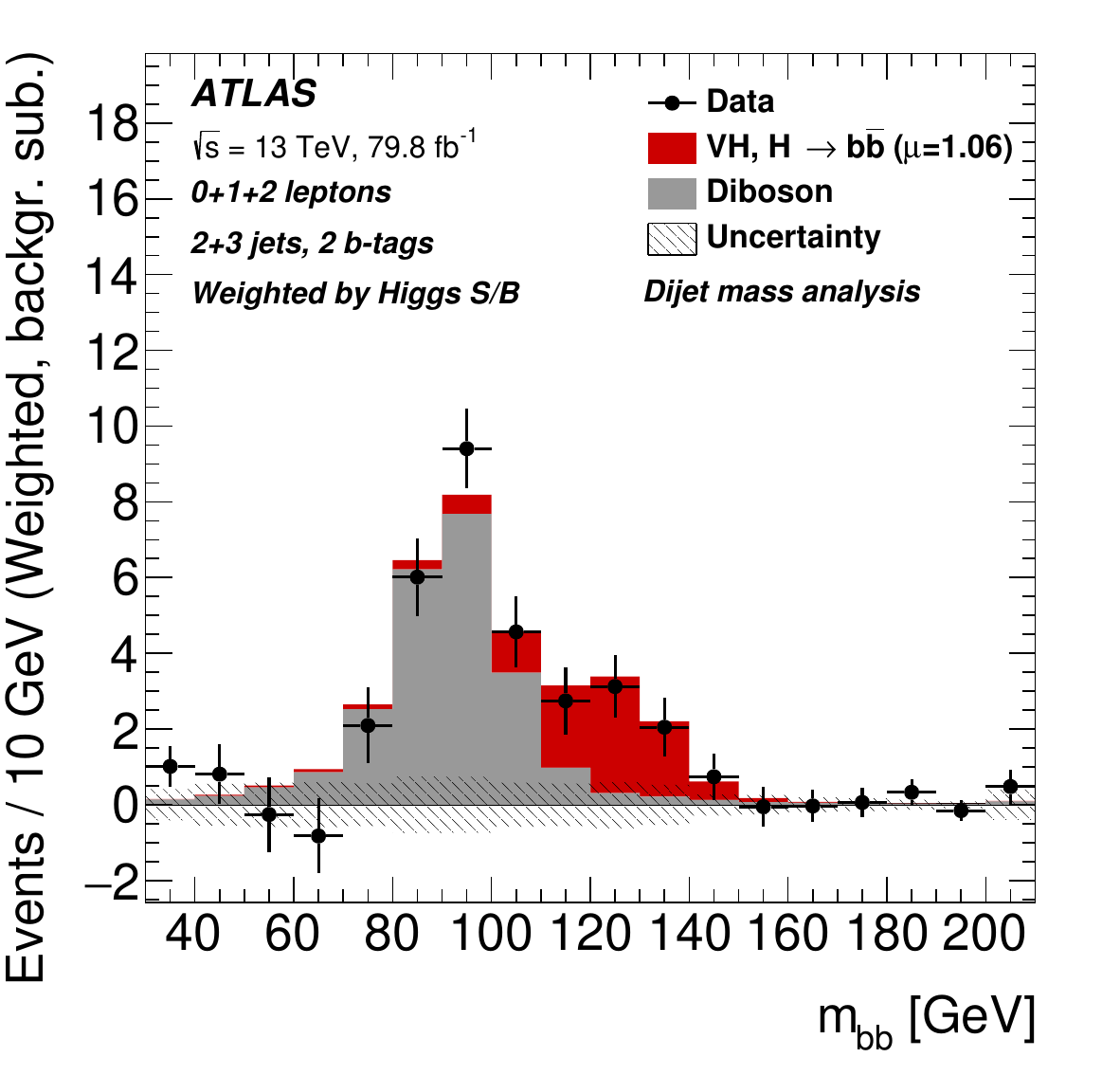}
\caption{
The invariant mass distribution of pairs of jets containing bottom quarks from the ATLAS experiment. All backgrounds except the diboson background have been subtracted. The contribution from the Higgs boson is shown in red and from the contribution from the diboson is shown in gray. All categories used in the analysis have been combined and weighted according to the signal-to-background ratio of each category. From Ref.\cite{ATLAS:2018kot}
\label{fig:HbbObs}
}
\end{figure}

The coupling of the Higgs to bottom quarks was observed in August 2018 by the ATLAS collaboration~\cite{ATLAS:2018kot}. It was confirmed by the CMS collaboration~\cite{CMS-PAS-HIG-18-016} shortly afterward. The value for $mu$ obtained was 
\begin{equation}
    \mu = 1.01 \pm 0.12 \mathrm{(stat.)} ^{+0.16}_{-0.15} \mathrm{(syst.)}
\end{equation}
which is highly consistent with the SM prediction. Figure~\ref{fig:HbbObs} shows the invariant mass distribution from this analysis after all backgrounds except that from the diboson have been subtracted. The data is shown by the black points and the fitted contribution for the Higgs boson signal is shown in red and the diboson background in gray. The events from the different categories used in the analysis have been combined and weighted according to the sensitivity of each channel otherwise the distribution would be dominated by those categories with high statistics but low signal sensitivity.

Already at the time of observation, the uncertainty was dominated by the systematic uncertainty. The largest uncertainties for the analysis at the time were:
\begin{itemize}
    \item Signal model: 0.09
    \item Monte Carlo statistics: 0.07
    \item V+jet normalization: 0.06
    \item b-tag: 0.06
\end{itemize}
The systematic uncertainty should not be viewed as necessarily the ultimate limit on the sensitivity of the analysis, because many of components of the systematic uncertainties can be expected to improve. For example, additional theoretical work would reduce uncertainties on the signal. Additional work to improve the performance of flavor tagging algorithms and better calibration uncertainties would reduce the b-tagging uncertainty. Finally, the impact of limited Monte Carlo statistics can be reduced by producing additional Monte Carlo (i.e. using additional computational resources) or by developing clever techniques to minimize their impact.

This observation confirms that the Higgs boson couples to bottom quarks and that the strength of the coupling is consistent within uncertainties with the predictions from the Standard Model.  Additional analyses have been performed since observation, studying the coupling to greater precision, studying the coupling strength in different categories and exploring the high momentum regime. To date, all results are consistent with the predictions of the Standard Model, but the full dataset from the LHC will increase the available statistics by an order of magnitude.

\section*{Acknowledgments}
I would like to thank Marjorie Shapiro, Simone Pagan Griso, Haichen Wang and Yury Kolomensky for material used to prepare these lectures. I would also like to thank Hadley Santana Quieroz for help in preparing this manuscript.

\nocite{*}
\bibliographystyle{plain}       
\bibliography{refs.bib}%

\begin{thebibliography}{10}

\bibitem{Aaij:2014zzy}
R.~Aaij et~al.
\newblock {Performance of the LHCb Vertex Locator}.
\newblock {\em JINST}, 9:P09007, 2014.

\bibitem{FASER:2022hcn}
Henso Abreu et~al.
\newblock {The FASER Detector}.
\newblock 7 2022.
\newblock arXiv:2207.11427.

\bibitem{MoEDAL:2014ttp}
B.~Acharya et~al.
\newblock {The Physics Programme Of The MoEDAL Experiment At The LHC}.
\newblock {\em Int. J. Mod. Phys. A}, 29:1430050, 2014.

\bibitem{LHCf:2008lfy}
O.~Adriani et~al.
\newblock {The LHCf detector at the CERN Large Hadron Collider}.
\newblock {\em JINST}, 3:S08006, 2008.

\bibitem{Aielli:2019ivi}
Giulio Aielli et~al.
\newblock {Expression of interest for the CODEX-b detector}.
\newblock {\em Eur. Phys. J. C}, 80(12):1177, 2020.

\bibitem{Alme:2010ke}
J.~Alme et~al.
\newblock {The ALICE TPC, a large 3-dimensional tracking device with fast
  readout for ultra-high multiplicity events}.
\newblock {\em Nucl. Instrum. Meth. A}, 622:316--367, 2010.

\bibitem{MATHUSLA:2018bqv}
Cristiano Alpigiani et~al.
\newblock {A Letter of Intent for MATHUSLA: A Dedicated Displaced Vertex
  Detector above ATLAS or CMS.}
\newblock 7 2018.
\newblock arXiv:1811.00927.

\bibitem{MATHUSLA:2020uve}
Cristiano Alpigiani et~al.
\newblock {An Update to the Letter of Intent for MATHUSLA: Search for
  Long-Lived Particles at the HL-LHC}.
\newblock 9 2020.
\newblock arXiv:2009.01693.

\bibitem{TOTEM:2008lue}
G.~Anelli et~al.
\newblock {The TOTEM experiment at the CERN Large Hadron Collider}.
\newblock {\em JINST}, 3:S08007, 2008.

\bibitem{BaileyLecture}
David Bailey.
\newblock Nuclear and particle physics, an introduction to subatomic physics,
  lecture 5.
\newblock
  \url{https://faraday.physics.utoronto.ca/PVB/DBailey/SubAtomic/Lectures/LectF05/Lect05.htm}.
\newblock Accessed: 2023-06-08.

\bibitem{Barney:2120661}
David Barney.
\newblock {CMS Detector Slice}.
\newblock \url{https://cds.cern.ch/record/2120661}, 2016.
\newblock CMS Collection.

\bibitem{BerryLectures}
Doug Berry.
\newblock Tracking detectors.
\newblock
  \url{https://indico.fnal.gov/event/54596/contributions/248563/attachments/158949/208802/Berry_Trackers_HCPSS22.pdf}.
\newblock Accessed: 2023-06-06.

\bibitem{BortolettoLecture1}
Daniela Bortoletto.
\newblock Detector technologies, lecture 1.
\newblock
  \url{https://indico.cern.ch/event/1023573/contributions/4400668/attachments/2300127/3912282/DB_HCPSS2021_Lect1.pdf
  }.
\newblock Accessed: 2023-06-06.

\bibitem{BortolettoLecture2}
Daniela Bortoletto.
\newblock Detector technologies, lecture 2.
\newblock
  \url{https://indico.cern.ch/event/1023573/contributions/4400678/attachments/2300862/3913720/DB_HCPSS2021_Lect2.pdf}.
\newblock Accessed: 2023-06-06.

\bibitem{Brice:1431477}
M~Brice.
\newblock {Images of the CMS ECAL Barrel (EB)}.
\newblock \url{https://cds.cern.ch/record/1431477}, 2008.
\newblock CMS Collection.

\bibitem{Maximilien:1211045}
Maximilien Brice.
\newblock {Views of the LHC tunnel in sector 3-4}.
\newblock \url{https://cds.cern.ch/record/1211045}, 2009.

\bibitem{Bruening:782076}
Oliver~Sim Bruening, Paul Collier, P~Lebrun, Stephen Myers, Ranko Ostojic, John
  Poole, and Paul Proudlock.
\newblock {LHC Design Report}.
\newblock Technical Report CERN-2004-003-V-1, CERN Yellow Reports: Monographs,
  Geneva, 2004.

\bibitem{Cacciari:2008gp}
Matteo Cacciari, Gavin~P. Salam, and Gregory Soyez.
\newblock {The anti-$k_t$ jet clustering algorithm}.
\newblock {\em JHEP}, 04:063, 2008.

\bibitem{Catani:1993hr}
S.~Catani, Yuri~L. Dokshitzer, M.~H. Seymour, and B.~R. Webber.
\newblock {Longitudinally invariant $K_t$ clustering algorithms for hadron
  hadron collisions}.
\newblock {\em Nucl. Phys. B}, 406:187--224, 1993.

\bibitem{CERN-EX-9308048_09}
CERN.
\newblock {ATLAS electromagnetic calorimeter layer}.
\newblock \url{https://cds.cern.ch/record/39737}, 1993.

\bibitem{Charpak:1968kd}
Georges Charpak, R.~Bouclier, T.~Bressani, J.~Favier, and C.~Zupancic.
\newblock {The Use of Multiwire Proportional Counters to Select and Localize
  Charged Particles}.
\newblock {\em Nucl. Instrum. Meth.}, 62:262--268, 1968.

\bibitem{CMS_DetectorPaper}
S.~Chatrchyan et~al.
\newblock {The CMS experiment at the CERN LHC}.
\newblock {\em JINST}, 3:S08004, 2008.

\bibitem{Chatrchyan:2012xdj}
Serguei Chatrchyan et~al.
\newblock {Observation of a new boson at a mass of 125 GeV with the CMS
  experiment at the LHC}.
\newblock {\em Phys. Lett. B}, 716:30--61, 2012.

\bibitem{ATLAS:1996guk}
ATLAS Collaboration.
\newblock {ATLAS calorimeter performance Technical Design Report}.
\newblock 12 1996.

\bibitem{ATLAS_DetectorPaper}
ATLAS Collaboration.
\newblock {The ATLAS Experiment at the CERN Large Hadron Collider}.
\newblock {\em JINST}, 3:S08003, 2008.

\bibitem{Aad:2012tfa}
ATLAS Collaboration.
\newblock {Observation of a new particle in the search for the Standard Model
  Higgs boson with the ATLAS detector at the LHC}.
\newblock {\em Phys. Lett. B}, 716:1--29, 2012.

\bibitem{ATLAS:2016nnj}
ATLAS Collaboration.
\newblock {Reconstruction of primary vertices at the ATLAS experiment in Run 1
  proton\textendash{}proton collisions at the LHC}.
\newblock {\em Eur. Phys. J. C}, 77(5):332, 2017.

\bibitem{ATLAS:2018kot}
ATLAS Collaboration.
\newblock {Observation of $H \rightarrow b\bar{b}$ decays and $VH$ production
  with the ATLAS detector}.
\newblock {\em Phys. Lett. B}, 786:59--86, 2018.

\bibitem{ATLAS:2021yvc}
ATLAS Collaboration.
\newblock {Expected tracking and related performance with the updated ATLAS
  Inner Tracker layout at the High-Luminosity LHC}.
\newblock 2021.

\bibitem{ATLAS:2020cli}
ATLAS Collaboration.
\newblock {Jet energy scale and resolution measured in
  proton\textendash{}proton collisions at $\sqrt{s}=13$~TeV with the ATLAS
  detector}.
\newblock {\em Eur. Phys. J. C}, 81(8):689, 2021.

\bibitem{ATLAS:2022vkf}
ATLAS Collaboration.
\newblock {A detailed map of Higgs boson interactions by the ATLAS experiment
  ten years after the discovery}.
\newblock {\em Nature}, 607(7917):52--59, 2022.
\newblock [Erratum: Nature 612, E24 (2022)].

\bibitem{ATLAS:2022qxm}
ATLAS Collaboration.
\newblock {ATLAS flavour-tagging algorithms for the LHC Run 2 $pp$ collision
  dataset}.
\newblock 11 2022.

\bibitem{cms:tracker}
CMS Collaboration.
\newblock Silicon strips.
\newblock \url{https://cms.cern/detector/identifying-tracks/silicon-strips}.
\newblock Accessed: 2023-03-28.

\bibitem{CMS:1997ema}
CMS Collaboration.
\newblock {CMS: The electromagnetic calorimeter. Technical design report}.
\newblock 12 1997.

\bibitem{CMS:2014pgm}
CMS Collaboration.
\newblock {Description and performance of track and primary-vertex
  reconstruction with the CMS tracker}.
\newblock {\em JINST}, 9(10):P10009, 2014.

\bibitem{CMS:2016lmd}
CMS Collaboration.
\newblock {Jet energy scale and resolution in the CMS experiment in pp
  collisions at 8 TeV}.
\newblock {\em JINST}, 12(02):P02014, 2017.

\bibitem{CMS:2017wtu}
CMS Collaboration.
\newblock {Identification of heavy-flavour jets with the CMS detector in pp
  collisions at 13 TeV}.
\newblock {\em JINST}, 13(05):P05011, 2018.

\bibitem{CMS-PAS-HIG-18-016}
CMS Collaboration.
\newblock {Observation of Higgs boson decay to bottom quarks}.
\newblock Technical report, CERN, Geneva, 2018.

\bibitem{CMS:2022dwd}
CMS Collaboration.
\newblock {A portrait of the Higgs boson by the CMS experiment ten years after
  the discovery}.
\newblock {\em Nature}, 607(7917):60--68, 2022.

\bibitem{ALICE_DetectorPaper}
The~ALICE Collaboration.
\newblock The {ALICE} experiment at the {CERN} {LHC}.
\newblock {\em \textnormal{2008} JINST}, 3:S08002.

\bibitem{LHCb_DetectorPaper}
The~LHCb Collaboration.
\newblock The {LHCb} detector at the {LHC}.
\newblock {\em \textnormal{2008} JINST}, 3:S08005.

\bibitem{lhcb:velo}
CERN Courier.
\newblock The last module of lhcb’s velo arrives.
\newblock
  \url{https://cerncourier.com/a/the-last-module-of-lhcbs-velo-arrives/}.
\newblock Accessed: 2023-03-28.

\bibitem{Dominguez}
Aaron Dominguez.
\newblock
  \url{https://indico.cern.ch/event/110193/contributions/1313715/attachments/32231/46718/New_Results_from_CMS_2010.pdf}.
\newblock Accessed: 2023-06-08.

\bibitem{Edwards:1992unz}
D.~A. Edwards and M.~J. Syphers.
\newblock {\em {An Introduction to the Physics of High-Energy Accelerators}}.
\newblock Wiley Series in Beam Physics and Accelerator Technology. Wiley, New
  York, 1992.

\bibitem{EldredLectures}
Jeffrey Eldred.
\newblock Accelerator physics concepts.
\newblock
  \url{https://indico.fnal.gov/event/54596/contributions/248603/attachments/159208/209196/JEldred\%20Collider\%20School.pdf}.
\newblock Accessed: 2023-06-05.

\bibitem{Ellis:1993tq}
Stephen~D. Ellis and Davison~E. Soper.
\newblock {Successive combination jet algorithm for hadron collisions}.
\newblock {\em Phys. Rev. D}, 48:3160--3166, 1993.

\bibitem{Haas:2014dda}
Andrew Haas, Christopher~S. Hill, Eder Izaguirre, and Itay Yavin.
\newblock {Looking for milli-charged particles with a new experiment at the
  LHC}.
\newblock {\em Phys. Lett. B}, 746:117--120, 2015.

\bibitem{KolbergLectures}
Ted Kolberg.
\newblock Calorimetry.
\newblock
  \url{https://indico.fnal.gov/event/54596/contributions/248562/attachments/158933/208776/tkolberg_calorimeters_hcpss22.pdf}.
\newblock Accessed: 2023-06-06.

\bibitem{Lopienska:2800984}
Ewa Lopienska.
\newblock {The CERN accelerator complex, layout in 2022. Complexe des
  accélérateurs du CERN en janvier 2022}.
\newblock 2022.
\newblock General Photo.

\bibitem{MangleLecture}
Stuart Mangles.
\newblock Introduction to plasma wakefield acceleration.
\newblock
  \url{https://indico.cern.ch/event/1192266/contributions/5012569/attachments/2543655/4379917/Wakefield_intro.pdf}.
\newblock Accessed: 2023-06-12.

\bibitem{sidet}
Vito Manzari.
\newblock Silicon detectors, lecture 2.
\newblock
  \url{https://indico.cern.ch/event/453690/sessions/99350/attachments/1184199/1726998/2015-11_SiliconDetectors_manzari_Lecture2.pdf}.
\newblock Accessed: 2023-03-31.

\bibitem{barn}
Mike Perricone.
\newblock Hitting the broad side of a (classified) barn.
\newblock {\em Symmetry}, 03, 2006.

\bibitem{PrebysAcceleratorCourse}
Eric Prebys.
\newblock Accelerator physics fundamentals online course.
\newblock
  \url{https://eprebys.faculty.ucdavis.edu/accelerator-physics-fundamentals-online-course/}.
\newblock Accessed: 2023-06-05.

\bibitem{shapiro}
Marjorie Shapiro.
\newblock personal communication.

\bibitem{RendeLecture1}
Rende Steerenberg.
\newblock Accelerator physics, lecture 1.
\newblock
  \url{https://indico.fnal.gov/event/43762/contributions/192686/attachments/132948/163699/rs20200813_HCPSS-2020_Lecture-1.pdf}.
\newblock Accessed: 2023-06-05.

\bibitem{RendeLecture2}
Rende Steerenberg.
\newblock Accelerator physics, lecture 2.
\newblock
  \url{https://indico.fnal.gov/event/43762/contributions/192695/attachments/133019/163819/rs20200814_HCPSS-2020_Lecture-2.pdf
  }.
\newblock Accessed: 2023-06-05.

\bibitem{Wigmans:2000vf}
R.~Wigmans.
\newblock {\em {Calorimetry: Energy measurement in particle physics}}, volume
  107.
\newblock 2000.

\bibitem{ParticleDataGroup:2022pth}
R.~L. Workman et~al.
\newblock {Review of Particle Physics}.
\newblock {\em PTEP}, 2022:083C01, 2022.

\end{thebibliography}

\end{document}